\documentclass[twocolumn]{aastex62}

\newcommand{\Ms}{$\rm M_{\odot}$}

\newcommand{\Mearth}{$\rm M_{\oplus}$} 
\newcommand{\Rearth}{$\rm R_{\oplus}$}
\newcommand{\degree}{$^{\circ}$}
\newcommand{\gcm}{$\rm g\ cm^{-3}$}

\usepackage{amsmath}
\usepackage{natbib}
\usepackage[flushleft]{threeparttable}
\usepackage[T1]{fontenc}
\usepackage{tipa}

\graphicspath{{./}{figures/}}

\received{}
\revised{}
\accepted{}
\submitjournal{ApJ}

\shorttitle{K2-146}
\shortauthors{Lam et al.}

\begin{document}
\title{It takes two planets in resonance to tango around K2-146}

\correspondingauthor{Kristine Lam}
\email{k.lam@tu-berlin.de}

\author{Kristine W. F. Lam}
\affiliation{Center for Astronomy and Astrophysics, Technical University Berlin, Hardenbergstr. 36, 10623 Berlin, Germany}

\author{Judith Korth}
\affiliation{Rheinisches Institut f\"ur Umweltforschung an der Universit\"at zu K\"oln, Aachener Strasse 209, D-50931 K\"oln Germany}

\author{Kento Masuda}
\affiliation{Department of Astrophysical Sciences, Princeton University, Princeton, NJ 08544, USA}
\affiliation{NASA Sagan Fellow}

\author{Szil\'ard Csizmadia}
\affiliation{Institute of Planetary Research, German Aerospace Center, Rutherfordstrasse 2, 12489 Berlin, Germany}

\author{Philipp Eigm\"uller}
\affiliation{Institute of Planetary Research, German Aerospace Center, Rutherfordstrasse 2, 12489 Berlin, Germany}

\author{Gu\dh mundur K\'ari Stef\'ansson}
\affiliation{Department of Astronomy and Astrophysics, The Pennsylvania State University, University Park, PA, USA}

\author{Michael Endl}
\affiliation{Department of Astronomy and McDonald Observatory, University of Texas at Austin, 2515 Speedway, Austin, TX 78712, USA}

\author{Simon Albrecht}
\affiliation{Stellar Astrophysics Centre, Department of Physics and Astronomy, Aarhus University, Ny Munkegade 120, DK-8000 Aarhus C, Denmark}

\author{Rafael Luque}
\affiliation{Instituto de Astrof\'\i sica de Canarias (IAC), 38205 La Laguna, Tenerife, Spain}
\affiliation{Departamento de Astrof\'\i sica, Universidad de La Laguna (ULL), 38206 La Laguna, Tenerife, Spain}

\author{John H. Livingston}
\affiliation{Department of Astronomy, The University of Tokyo, 7-3-1 Hongo, Bunkyo-ku, Tokyo 113-0033, Japan}

\author{Teruyuki Hirano}
\affiliation{Department of Earth and Planetary Sciences, Tokyo Institute of Technology, 2-12-1 Ookayama, Meguro-ku, Tokyo, Japan}

\author{Roi Alonso~Sobrino}
\affiliation{Instituto de Astrof\'\i sica de Canarias (IAC), 38205 La Laguna, Tenerife, Spain}
\affiliation{Departamento de Astrof\'\i sica, Universidad de La Laguna (ULL), 38206 La Laguna, Tenerife, Spain}

\author{Oscar Barrag\'{a}n}
\affiliation{Sub-department of Astrophysics, Department of Physics, University of Oxford, Oxford OX1 3RH, UK}

\author{Juan Cabrera}
\affiliation{Institute of Planetary Research, German Aerospace Center (DLR), Rutherfordstrasse 2, D-12489 Berlin, Germany}

\author{Ilaria Carleo}
\affiliation{Astronomy Department and Van Vleck Observatory, Wesleyan University, Middletown, CT 06459, USA}

\author{Alexander Chaushev}
\affiliation{Center for Astronomy and Astrophysics, TU Berlin, Hardenbergstr. 36, 10623 Berlin, Germany}

\author{William D. Cochran}
\affiliation{Department of Astronomy and McDonald Observatory, University of Texas at Austin, 2515 Speedway, Austin, TX 78712, USA}

\author{Fei Dai}
\affiliation{Department of Astrophysical Sciences, Princeton University, 4 Ivy Lane, Princeton, NJ, 08544, USA}
\affiliation{Department of Physics and Kavli Institute for Astrophysics and Space Research, MIT, Cambridge, MA 02139, USA}

\author{Jerome de Leon}
\affiliation{Department of Astronomy, The University of Tokyo, 7-3-1 Hongo, Bunkyo-ku, Tokyo 113-0033, Japan}

\author{Hans J. Deeg}
\affiliation{Instituto de Astrof\'\i sica de Canarias (IAC), 38205 La Laguna, Tenerife, Spain}
\affiliation{Departamento de Astrof\'\i sica, Universidad de La Laguna (ULL), 38206 La Laguna, Tenerife, Spain}

\author{Anders Erikson}
\affiliation{Institute of Planetary Research, German Aerospace Center (DLR), Rutherfordstrasse 2, D-12489 Berlin, Germany}

\author{Massimiliano Esposito}
\affiliation{Th\"uringer Landessternwarte Tautenburg,  D-07778 Tautenburg, Germany}

\author{Malcolm Fridlund}
\affiliation{Chalmers University of Technology, Department of Space, Earth and Environment, Onsala Space Observatory,  SE-439 92 Onsala, Sweden}
\affiliation{Leiden Observatory, University of Leiden, PO Box 9513, 2300 RA, Leiden, The Netherlands}

\author{Akihiko Fukui}
\affiliation{National Astronomical Observatory of Japan, NINS, 2-21-1 Osawa, Mitaka, Tokyo 181-8588, Japan}

\author{Davide Gandolfi}
\affiliation{Dipartimento di Fisica, Universit\'a di Torino, Via P. Giuria 1, I-10125, Torino, Italy}

\author{Iskra Georgieva}
\affiliation{Chalmers University of Technology, Department of Space, Earth and Environment, Onsala Space Observatory,  SE-439 92 Onsala, Sweden}

\author{Luc\'a Gonzalez Cuesta}
\affiliation{Instituto de Astrof\'\i sica de Canarias (IAC), 38205 La Laguna, Tenerife, Spain}
\affiliation{Departamento de Astrof\'\i sica, Universidad de La Laguna (ULL), 38206 La Laguna, Tenerife, Spain}

\author{Sascha Grziwa}
\affiliation{Rheinisches Institut f\"ur Umweltforschung an der Universit\"at zu K\"oln, Aachener Strasse 209, D-50931 K\"oln Germany}

\author{Eike W. Guenther}
\affiliation{Th\"uringer Landessternwarte Tautenburg,  D-07778 Tautenburg, Germany}

\author{Artie P. Hatzes}
\affiliation{Th\"uringer Landessternwarte Tautenburg,  D-07778 Tautenburg, Germany}

\author{Diego Hidalgo}
\affiliation{Instituto de Astrof\'\i sica de Canarias (IAC), 38205 La Laguna, Tenerife, Spain}
\affiliation{Departamento de Astrof\'\i sica, Universidad de La Laguna (ULL), 38206 La Laguna, Tenerife, Spain}

\author{Maria Hjorth}
\affiliation{Stellar Astrophysics Centre, Department of Physics and Astronomy, Aarhus University, Ny Munkegade 120, DK-8000 Aarhus C, Denmark}

\author{Petr Kabath}
\affiliation{Astronomical Institute, Czech Academy of Sciences, Fri\v{c}ova 298, 25165, Ond\v{r}ejov, Czech Republic}

\author{Emil Knudstrup}
\affiliation{Stellar Astrophysics Centre, Department of Physics and Astronomy, Aarhus University, Ny Munkegade 120, DK-8000 Aarhus C, Denmark}

\author{Mikkel N. Lund}
\affiliation{Stellar Astrophysics Centre, Department of Physics and Astronomy, Aarhus University, Ny Munkegade 120, DK-8000 Aarhus C, Denmark}

\author{Suvrath Mahadevan}
\affiliation{Department of Astronomy and Astrophysics, The Pennsylvania State University, 525 Davey Lab, University Park, PA 16802, USA}
\affiliation{Center for Exoplanets and Habitable Worlds, University Park, PA 16802, USA}
\affiliation{Penn State Astrobiology Research Center, University Park, PA 16802, USA}

\author{Savita Mathur}
\affiliation{Instituto de Astrof\'\i sica de Canarias (IAC), 38205 La Laguna, Tenerife, Spain}
\affiliation{Departamento de Astrof\'\i sica, Universidad de La Laguna (ULL), 38206 La Laguna, Tenerife, Spain}

\author{Pilar Monta\~nes Rodr\'iguez}
\affiliation{Instituto de Astrof\'\i sica de Canarias (IAC), 38205 La Laguna, Tenerife, Spain}
\affiliation{Departamento de Astrof\'\i sica, Universidad de La Laguna (ULL), 38206 La Laguna, Tenerife, Spain}

\author{Felipe Murgas}
\affiliation{Instituto de Astrof\'\i sica de Canarias (IAC), 38205 La Laguna, Tenerife, Spain}
\affiliation{Departamento de Astrof\'\i sica, Universidad de La Laguna (ULL), 38206 La Laguna, Tenerife, Spain}

\author{Norio Narita}
\affiliation{Astrobiology Center, NINS, 2-21-1 Osawa, Mitaka, Tokyo 181-8588, Japan}
\affiliation{JST, PRESTO, 2-21-1 Osawa, Mitaka, Tokyo 181-8588, Japan}
\affiliation{National Astronomical Observatory of Japan, NINS, 2-21-1 Osawa, Mitaka, Tokyo 181-8588, Japan}
\affiliation{Instituto de Astrof\'\i sica de Canarias (IAC), 38205 La Laguna, Tenerife, Spain}

\author{David Nespral}
\affiliation{Instituto de Astrof\'\i sica de Canarias (IAC), 38205 La Laguna, Tenerife, Spain}
\affiliation{Departamento de Astrof\'\i sica, Universidad de La Laguna (ULL), 38206 La Laguna, Tenerife, Spain}

\author{Prajwal Niraula}
\affiliation{Department of Earth, Atmospheric and Planetary Sciences, MIT, 77 Massachusetts Avenue, Cambridge, MA 02139, USA}

\author{Enric Palle}
\affiliation{Instituto de Astrof\'\i sica de Canarias (IAC), 38205 La Laguna, Tenerife, Spain}
\affiliation{Departamento de Astrof\'\i sica, Universidad de La Laguna (ULL), 38206 La Laguna, Tenerife, Spain}

\author{Martin~P\"atzold}
\affiliation{Rheinisches Institut f\"ur Umweltforschung an der Universit\"at zu K\"oln, Aachener Strasse 209, D-50931 K\"oln Germany}

\author{Carina M. Persson}
\affiliation{Chalmers University of Technology, Department of Space, Earth and Environment, Onsala Space Observatory,  SE-439 92 Onsala, Sweden}

\author{Jorge Prieto-Arranz}
\affiliation{Instituto de Astrof\'\i sica de Canarias (IAC), 38205 La Laguna, Tenerife, Spain}
\affiliation{Departamento de Astrof\'\i sica, Universidad de La Laguna (ULL), 38206 La Laguna, Tenerife, Spain}

\author{Heike Rauer}
\affiliation{Institute of Planetary Research, German Aerospace Center (DLR), Rutherfordstrasse 2, D-12489 Berlin, Germany}
\affiliation{Center for Astronomy and Astrophysics, TU Berlin, Hardenbergstr. 36, 10623 Berlin, Germany}
\affiliation{Institute of Geological Sciences, FU Berlin, Malteserstr. 74-100, D-12249 Berlin}

\author{Seth Redfield}
\affiliation{Astronomy Department and Van Vleck Observatory, Wesleyan University, Middletown, CT 06459, USA}

\author{Ignasi Ribas}
\affiliation{Institut de Ci\`encies de l'Espai (ICE, CSIC), Campus UAB,C/ de Can Magrans s/n, E-08193 Bellaterra, Spain}
\affiliation{Institut d'Estudis Espacials de Catalunya (IEEC), C/ Gran Capit\`a 2-4, E-08034 Barcelona, Spain}

\author{Paul Robertson}
\affiliation{Department of Astronomy and Astrophysics, The Pennsylvania State University, University Park, PA, USA}

\author{Marek Skarka}
\affiliation{Astronomical Institute, Czech Academy of Sciences, Fri\v{c}ova 298, 25165, Ond\v{r}ejov, Czech Republic}
\affiliation{Department of Theoretical Physics and Astrophysics, Masaryk University, Kotl\'{a}\v{r}sk\'{a} 2, 61137 Brno, Czech Republic}

\author{Alexis M. S. Smith}
\affiliation{Institute of Planetary Research, German Aerospace Center (DLR), Rutherfordstrasse 2, D-12489 Berlin, Germany}

\author{Jan	Subjak}
\affiliation{Astronomical Institute, Czech Academy of Sciences, Fri\v{c}ova 298, 25165, Ond\v{r}ejov, Czech Republic}
\affiliation{Astronomical Institute, Faculty of Mathematics and Physics, Charles University, Ke Karlovu 2027/3, 12116 Prague, Czech Republic}

\author{Vincent Van Eylen}
\affiliation{Department of Astrophysical Sciences, Princeton University, 4 Ivy Lane, Princeton, NJ, 08544, USA}



\begin{abstract}
K2-146 is a cool, 0.358\Ms\ dwarf that was found to host a mini-Neptune with a 2.67-days period. The planet exhibited strong transit timing variations (TTVs) of greater than 30 minutes, indicative of the presence of a further object in the system. Here we report the discovery of the previously undetected outer planet, K2-146 c, in the system using additional photometric data. K2-146 c was found to have a grazing transit geometry and a 3.97-day period. The outer planet was only significantly detected in the latter \textit{K2} campaigns presumably because of precession of its orbital plane. The TTVs of K2-146 b and c were measured using observations spanning a baseline of almost 1200 days. We found strong anti-correlation in the TTVs, suggesting the two planets are gravitationally interacting. Our TTV and transit model analyses revealed that K2-146 b has a radius of 2.25 $\pm$ 0.10 \Rearth\ and a mass of 5.6 $\pm$ 0.7 \Mearth, whereas K2-146 c has a radius of $2.59_{-0.39}^{+1.81}$ \Rearth\ and a mass of 7.1 $\pm$ 0.9 \Mearth. The inner and outer planets likely have moderate eccentricities of $e = 0.14 \pm 0.07$ and $0.16 \pm 0.07$, respectively. Long-term numerical integrations of the two-planet orbital solution show that it can be dynamically stable for at least 2 Myr. The evaluation of the resonance angles of the planet pair indicates that K2-146 b and c are likely trapped in a 3:2 mean motion resonance. The orbital architecture of the system points to a possible convergent migration origin.

\end{abstract}



\section{Introduction} \label{sec:intro}

The \textit{Kepler} \citep{2010Sci...327..977B} and \textit{K2} \citep{2014PASP..126..398H} missions have brought about many exciting discoveries since the spacecraft was launched in 2009. Statistical studies using the \textit{Kepler} planet sample revealed that sub-Neptune size planets with $R_p < 4$ \Rearth~ are by far the most common type of planets in the galaxy (e.g. \cite{2011ApJ...728..117B}; \cite{2012ApJS..201...15H}; \cite{2013ApJS..204...24B}; \cite{2013ApJ...767...95D}; \cite{2013PNAS..11019273P}; \cite{2013ApJ...766...81F}). Previous works have also shown that short-period planets with radii between 1.5-6 \Rearth\ are common in near co-planar multi-planet systems \citep{2011ApJS..197....8L,2014ApJ...790..146F}. More recently, \cite{2018AJ....155...48W} used precisely determined stellar and planetary parameters to show that, multi-planet systems are dynamically packed and that adjacent planets in the same system are likely to have similar sizes.

Multi-planet systems are of particular interest because these systems can provide insights on the formation and evolution of our own Solar system. Gaining more knowledge about these systems is important for understanding the planetary system dynamics. To address these topics we need to know the planetary parameters, in particular their radii and masses. Unfortunately, only a small number of planets with precisely measured masses are known because there are inadequate telescope resources for sufficient spectroscopic measurements, or the stars are simply too faint. The mass determination for Earth- or Neptune-sized planets around Sun-like stars or faint stars by radial velocity (RV) is particularly difficult with currently available telescopes and instrumental technique because of the small Doppler reflex motion of the host star.

In multi-planet systems, planets can experience mutual gravitational interactions that perturb their orbits. One of the consequences of these is that individual transits vary periodically around a mean orbital period. This effect is referred to Transit Timing Variations  \citep[TTVs, e.g.\ ][]{holman_murray_2005,agol_et_al_2005}. This effect is most prominent when the orbital periods of the planets are close to a Mean-Motion Resonance (MMR), and it can be measured even for low mass planets. Thus TTVs are sometimes the only chance to characterize the planetary system. For example, a TTV analysis of KOI-142 revealed a pair of planets orbiting in a near 2:1 resonance \citep{2013ApJ...777....3N}. Extensive RV observations were obtained for K2-19 b and c, a two planet system in a near 3:2 MMR \citep{2015A&A...582A..33A,2015ApJ...815...47N,2015MNRAS.454.4267B,2017A&A...601A.128N}. The measured RV masses of the Neptune-sized planets were found to be consistent with the TTV-derived masses. Precise mass determination via TTVs showed a pair of planets, Kepler-36 b and c, that have distinctly different bulk densities, hinting at different formation origin of the planets \citep{2012Sci...337..556C}. Dynamical modelling of both the TTV and transit duration variations (TDVs) can also reveal mutual inclination of a pair of planets  \citep[e.g.\ Kepler-108; ][]{2017AJ....153...45M} and uncover the presence of an additional non-transiting companion in some cases (e.g. KOI-872 system; \citealt{2012Sci...336.1133N}, Kepler-448 b and Kepler-693 b; \citealt{2017AJ....154...64M}). 

\subsection{The cool star K2-146}
K2-146 was first observed in the \textit{K2} Campaign 5. The $\sim 2.2R_{\oplus}$ mini-Neptune, K2-146 b, was validated independently by \cite{2018AJ....155..127H} and \cite{2018AJ....156..277L}. K2-146 b orbits around an M3.0V dwarf and was reported to have an orbital period of $2.645$~days.
The system was also independently flagged as a planetary candidate by \cite{2016MNRAS.461.3399P}, \cite{2016MNRAS.463.1780L}, and \cite{2017AJ....154..207D}. \citeauthor{2018AJ....155..127H} reported strong TTVs with amplitude of over 30 minutes. The observed orbital perturbation of the planet is likely caused by either a massive object in its vicinity or an additional object orbiting in or close to a MMR. The stellar parameters of K2-146 are summarised in Table \ref{tab:stellarparam}.

This paper is organised as follows. Section \ref{sec:K2photometry} describes the \textit{K2} photometric observations, data reduction and planet detection. Section \ref{sec:timing_extraction} describes the extraction of transit times for K2-146 b and c. The TTV model and analysis are described in Section \ref{sec:TTV_analysis}. Subsequent derivation of transit parameters of the mini-Neptune pair are presented in Section \ref{sec:tlcm_analysis}. In Section \ref{sec:discussion} we evaluate the stability, orbital resonance, and interior composition of the planets pair, and interpret their possible implications on the evolution history of the system. Finally, we draw our conclusions in Section.
\ref{sec:conclusion}.

\begin{table}[htbp!]
\begin{threeparttable}
\caption{Stellar parameters and photometric magnitudes of K2-146. \label{tab:stellarparam}}
\begin{tabular}{lcc}
\toprule
Parameter                  & Value and uncertainty & Source \\
\hline
EPIC                       &    211924657         &     a     \\
2MASS                      &2MASS J08400641+1905346&    b     \\
Gaia                       & 661192902209491456   &     c     \\
RA                         &    08 40 06.42       &     c     \\
DEC                        &    +19 05 34.42      &     c     \\
$\mu_{RA}$ $\rm [mas/yr]$  &   $-15.92 \pm 0.12$  &     c     \\
$\mu_{Dec}$ $\rm [mas/yr]$ &   $-129.02 \pm 0.07$ &     c     \\
Parallax $\rm [mas]$       &   $12.582 \pm 0.075$ &     c     \\
Spectral type              &        M3.0V         &     d     \\
$T_{\rm eff}$ $\rm [K]$    &   $3385 \pm 70$      &     d     \\
$\rm [Fe/H]$ $[dex]$       &   $-0.02 \pm 0.12$   &     d     \\
$\log g$                   &   $4.906 \pm 0.041$  &     d     \\
Ms $[M_{\odot}]$           &  $0.358 \pm 0.042$   &     d     \\
Rs $[R_{\odot}]$           &  $0.350 \pm 0.035$   &     d     \\
Ls $[L_{\odot}]$           &  $0.015 \pm 0.003$   &     d     \\
\hline
\multicolumn{3}{l}{\textit{Photometric magnitudes}}	\\
Kep                        &        15.03         &     a     \\
Gaia G                     &        14.98         &     c     \\
Johnson B                  &        17.69         &     e     \\
Johnson V                  &        16.18         &     e     \\
J                          &        12.18         &     b     \\
H                          &        11.60         &     b     \\
K                          &        11.37         &     b     \\
\hline      \\

\end{tabular}
\begin{tablenotes}
\small
\item References of sources: (a) EXOFOP-K2: \url{https://exofop.ipac.caltech.edu/k2/}; (b) The Two Micron All Sky Survey (2MASS; \cite{2006AJ....131.1163S}); (c) Gaia DR2 \citep{2018A&A...616A...1G,2016A&A...595A...1G}; (d) \cite{2018AJ....155..127H}; (e) The AAVSO Photometric All-Sky Survey (APASS; \citealt{2009AAS...21440702H})
\end{tablenotes}

\end{threeparttable}
\end{table}

\section{\textit{K2} photometry \label{sec:K2photometry}} 
K2-146 was observed in Campaign 5, 16 and 18 (hereafter, C05, C16, and C18, respectively) in the long cadence mode. The photometric observations were obtained between 2015 April 27 and 2018 July 02, spanning a baseline of almost 1200 days. The \textit{K2} target pixel data was downloaded from the Mikulski Archive for Space Telescope\footnote{\url{http://archive.stsci.edu/kepler/data_search/search.php}} (MAST). A custom pipeline was implemented for light curve reduction and is described below. 

The photometric analysis was conducted for each campaign separately. For each campaign, the timestamps were combined and the quality of the light curve was tested using different thresholds to the number of counts per pixel. An optimal aperture was selected using the 100 counts per pixel threshold. Using this aperture we calculated the flux for each frame. To correct for possible correlation with the movement of pointing, we cut the light curves into segments with a length of 4 days. Between adjacent segments, there is an overlapping region of 0.8 days. The overlapping regions help us to avoid edge effects when fitting the data in the time domain to remove stellar variability. 

For each light curve segment, outliers (including transits) were identified and masked before fitting a multidimensional polynomial, over the POS\_CORR columns (which measures the relative motion of the star), to the data. To avoid influence from time-dependant variability, we simultaneously obtained a third order polynomial fit over the time. This fit was then applied to the whole segment, including the transits, to correct for possible correlation with the telescope pointing. After the correlation to the POS\_CORR columns have been removed, we masked outliers and transit events again before fitting a seventh order polynomial to the light curve segments to correct for stellar variability. Finally, all light curve segments are normalized and stacked together.

\begin{figure*}[htbp!]
\begin{center}
\includegraphics[width=\textwidth]{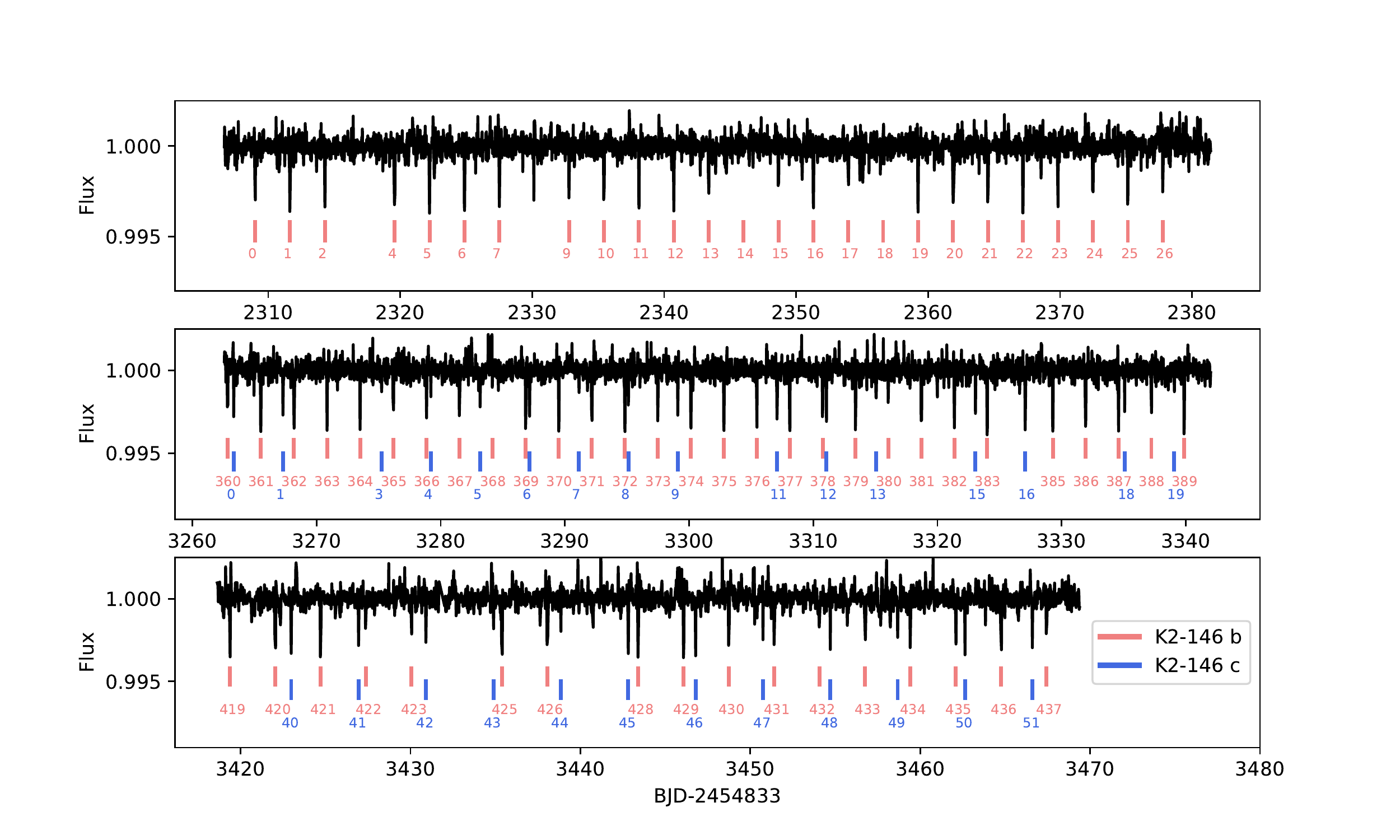}
\caption{\textit{K2} light curve of K2-146 from Campaigns 5 (top panel), Campaign 16 (middle panel) and Campaign 18 (bottom panel). The red and blue lines indicate transits of K2-146 b and K2-146 c used in transit modelling and TTV analysis. The numbers below the lines correspond to the integer transit epoch number from when the first transit became visible.  \label{fig:K2_lightcurve}}
\end{center}
\end{figure*}

We compared light curves generated from our custom pipeline with ones that are publicly available from the \cite{2014PASP..126..948V} pipeline (K2SFF) and from \cite{2018AJ....156...99L} (\texttt{EVEREST}; kindly provided by Luger). We found that the noise level of the light curves from the three pipelines are comparable. Light curves from Campaign 16 and Campaign 18 generated from our pipeline have a slightly lower overall scatter, whereas the scatter in the Campaign 5 lightcurve is slightly higher than the K2SFF pipeline. For a consistent analysis, we opted to use the light curves obtained from our pipeline (as shown in Figure \ref{fig:K2_lightcurve}) for light curve modelling and TTV analysis. 

\subsection{Planet detection}
\begin{figure}[htbp!]
\begin{center}
\includegraphics[trim= 30 10 0 20 ,width=0.5\textwidth]{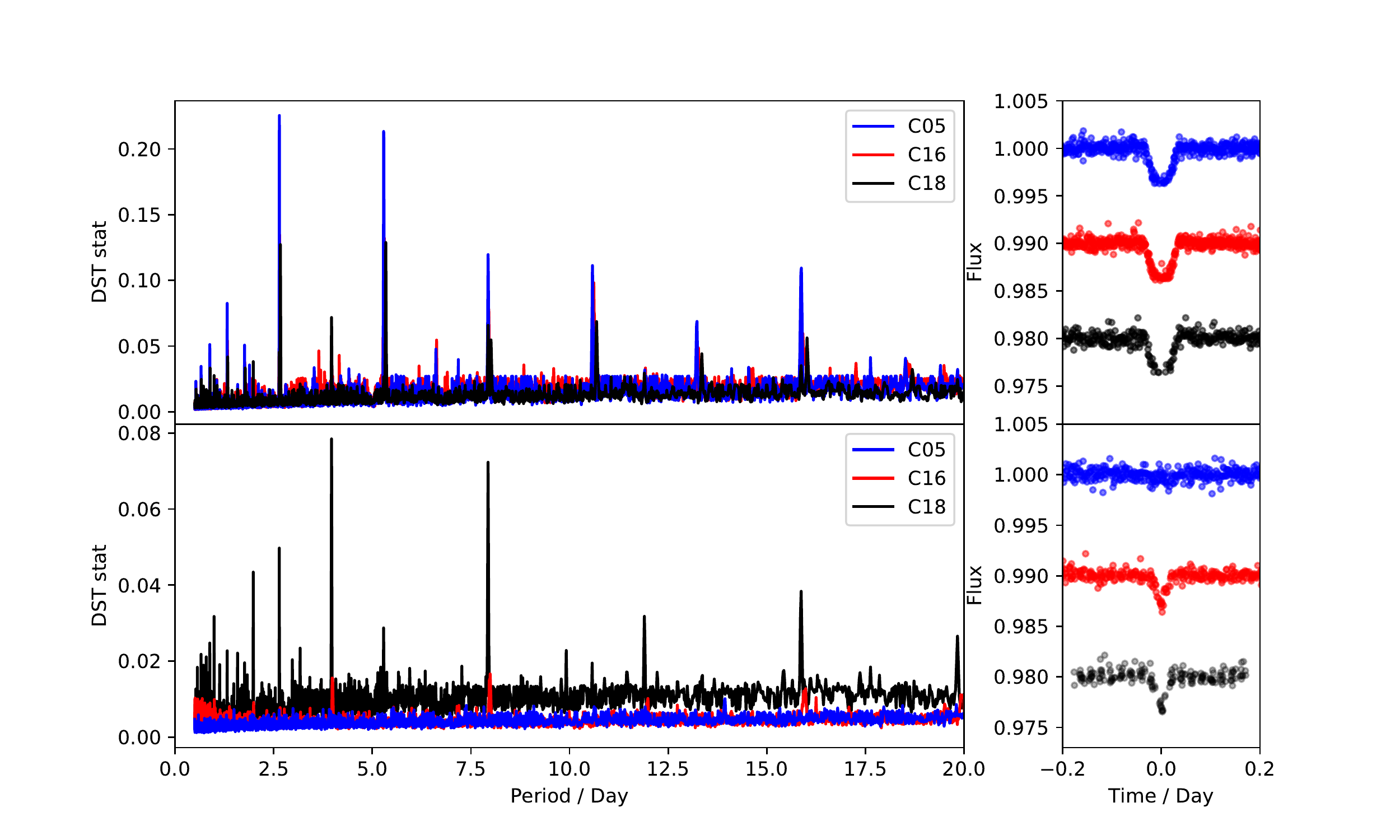}
\caption{\textit{Top left}: Periodogram of the DST statistics evaluated as in \cite{2012A&A...548A..44C} for K2-146. We detect K2-146 b with the strongest peak at $\sim2.6$ d. \textit{Top right}: Phase-folded light curves of K2-146 with TTV correction. The light curves are arbitrarily shifted for clarity. \textit{Bottom left}: Periodogram of the DST statistics after the signal of the inner planet is filtered. \textit{Bottom right}: Filtered, phase-folded light curves of K2-146 with TTV correction. The light curves are arbitrarily shifted for clarity. Transits of the outer planet were not significantly detected in C05, we attribute this effect to possible nodal precession of the orbital planet. The C05 light curve is phase-folded with the ephemeris derived from our transit search algorithm. \label{fig:dst_stat}}
\end{center}
\end{figure}

We searched the \textit{K2} light curves for transit signals using the \texttt{DST} algorithm \citep{2012A&A...548A..44C}, which optimizes the fit to the transit shapes with a parabolic function. Figure \ref{fig:dst_stat} shows the periodograms of the DST statistics measured in all light curves. The top left panel of Figure \ref{fig:dst_stat} shows that the $\sim2.6$ days signal of K2-146 b was detected in the C05, and subsequently in C16 and C18. 

The 2.6-day signal was then filtered and the light curves were analyzed with the \texttt{DST} algorithm again. Strong peaks at $\sim4$ days are found in the periodograms of the C16 and C18 data, as shown in the bottom left panel of Figure \ref{fig:dst_stat}. However, no significant detection is found in C05, and transits of the outer planet were not observed upon visual inspection due to the noise level of the C05 light curve. We also ran our transit search algorithm on the \texttt{K2SFF} C05 data since it has a slightly lower scatter. Although we detected hints of transit signal at $\sim4$ days, the detection was not significant. We attribute this to a precessing orbital plane of this outer planet, which we discuss in later sections.

The characterization of the multi-planet system K2-146 follows the approach outlined here: The transit parameters are derived iteratively. We first performed a global analysis to extract the transit times and transit parameters of planet b and planet c (Section \ref{sec:timing_extraction}). The transit parameters of the two planets were analysed independently using a stacked transit light curve. 
We then model the transit times of the planets to derive their respective orbital elements (Section \ref{sec:TTV_analysis}). Finally, we use information from the TTV-deduced orbital elements to model the stacked transits of planet b and planet c, and improve the precision of the system parameters (Section \ref{sec:tlcm_analysis}).

\section{Transit time extraction \label{sec:timing_extraction}} 

\subsection{PyTV \label{sec:pytv}} 
    We extracted the transit times using the Python Tool for Transit Variations (\texttt{PyTV}, Korth 2019, \textit{in prep.}). This tool uses \texttt{PyTransit} \citep{2015MNRAS.450.3233P} for transit modelling, \texttt{PyDE}\footnote{ \url{https://github.com/hpparvi/PyDE}} for optimisation and \texttt{emcee} \citep{2013PASP..125..306F} for posterior sampling. 
    
    Before modelling, the light curve around each transit were detrended by subtracting a second order polynomial fit to the out-of-transit light curve. These cut-out segments were then the input for the detailed modelling with \texttt{PyTV}. Some transits were excluded from the analysis (see Figure \ref{fig:K2_lightcurve}): transit number 3, 8, 384, 424 and 427 for planet b and transit numbers 2, 10, 14 and 17 for planet c.\footnote{We checked the light curve flags and found that transit number 3 and 424 both have flags 2048 (impulsive outlier) and 1048576 (thruster firing), transit number 8 has flag 8192 (cosmic ray), transit number 384 has flag 1024 (sudden sensitivity dropout), and transit number 427 has flag 32768 (no fine point). For planet c the transits with numbers 2 and 10 have flag 1048576. For transit number 14 and 17 we found no non-zero flags but they might not be visible because of the long cadence observations (30 minutes) and are therefore missed in the light curve.}

    The transit time extraction and transit fits were performed for each planet independently. The individual transits are fitted collectively with the \cite{2002ApJ...580L.171M} model, each with their own transit center but sharing the rest of the transit parameters (individual transit fitting models are found in appendix Figures \ref{fig:lc_individual_b} and \ref{fig:lc_individual_c}). The optimization was done by computing the log-likelihood, $\log L$, in our code to estimate transit parameters:
    \begin{equation}
        \log L = -\frac{1}{2} \sum_{i=0}^{N} \left[ \frac{(x_i-\mu_i)^2}{\sigma_i^2} + \ln{(2\pi \sigma_i^2)} \right]\,\, ,
        \label{eq:Log}
    \end{equation}
    where $x_i$ is the model, $\mu_i$ is the data, and $\sigma_i$ is the error in the data for the $i^{\rm th}$ point, respectively. The fitted parameters are radius ratio $k$, impact parameter $b$, stellar density $\rho_s$, all with uniform priors. We used a quadratic limb darkening model with the `triangular sampling’ parameterization presented by \cite{2013MNRAS.435.2152K}. Gaussian priors were imposed on the limb darkening coefficients, $u_1$ and $u_2$, where the central values of the coefficients were calculated with \texttt{PyLDTK} \citep{2015MNRAS.453.3821P} which utilises the spectrum library of \cite{2013A&A...553A...6H}. The transits of planet c is grazing, the $a/R_s$ and hence the mean stellar density is less well constraint. Thus the stellar density posterior from the planet b analysis was used as a stellar density prior in the planet c analysis. To account for the long exposure times of the \textit{K2} observation we applied a supersampling (n=10) as suggested in \cite{kipping_2010c}.
    
    For posterior sampling, we ran 5 MCMC chains with 5000 steps whereby the previous run was used as a burn-in for the current MCMC run. The chains were checked for convergence visually. The posteriors and the derived planetary parameters are shown in appendix Table \ref{tab:transitparams}. The resulting corner plots are shown in the appendix (Figures \ref{fig:lc_cornerplot_b} and \ref{fig:lc_cornerplot_c}). The transit times of both planets are used for a detailed TTV analysis in the following section. 
    
\begin{figure*}[htbp!]
\begin{center}
\includegraphics[width=\textwidth]{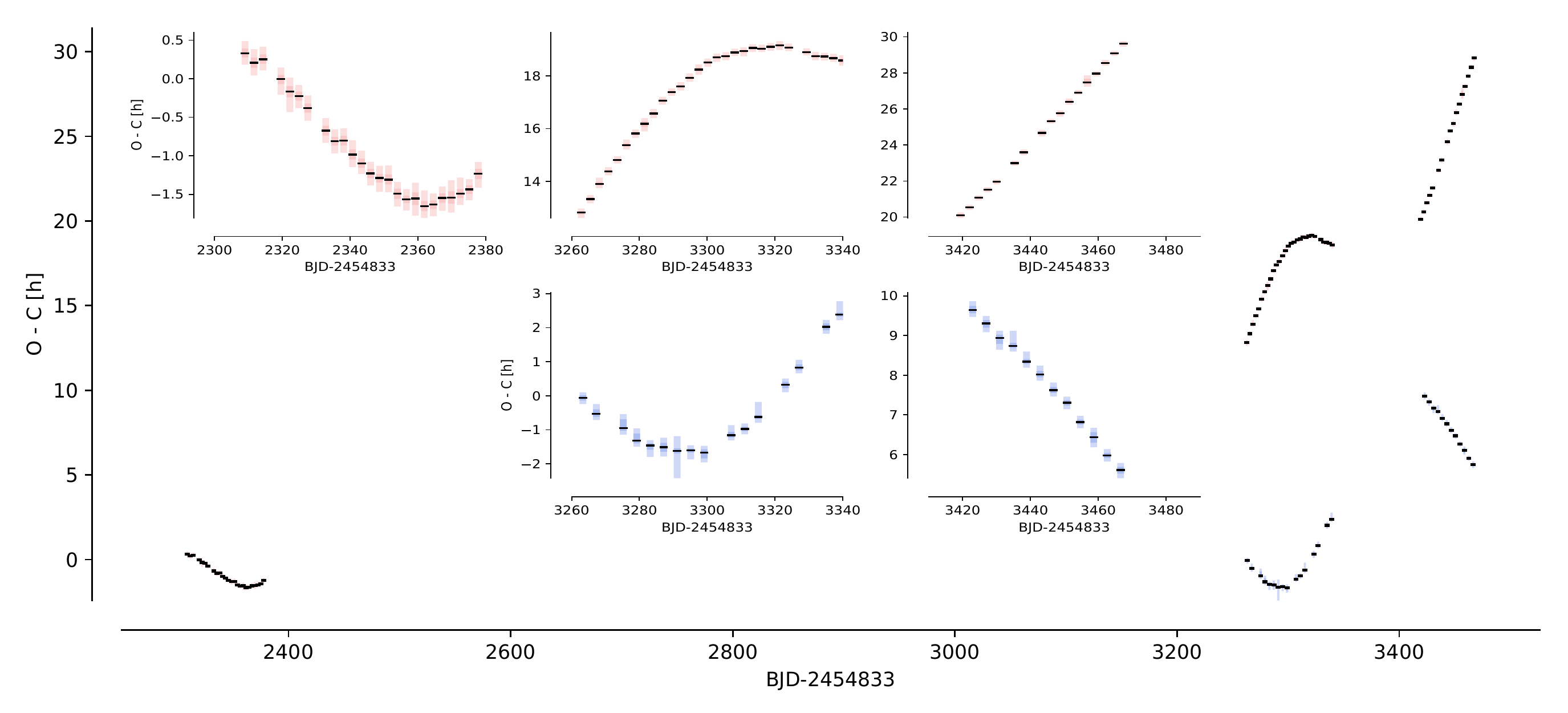}
\caption{ O$-$C diagram of planet b (red) and planet c (blue). An anti-correlation in transit times is clearly visible in C16 and C18. The inset shows a magnification of individual campaigns. The coloured error bars mark the percentiles of the fitted transit times. \label{fig:O-C}}
\end{center}
\end{figure*}

    The fitted transit times of planet b (red) and planet c (blue) are shown in the O$-$C diagram in figure \ref{fig:O-C}. An anti-correlation between the O$-$C values is clearly visible in campaigns 16 and 18. This shows that the two planets are orbiting in the same system and that they are gravitationally interacting with each other, confirming the planetary nature of the signals. For a better visualisation, the inset in figure \ref{fig:O-C} shows a magnification of the different campaigns. The coloured error bars mark the 1-$\sigma$ and 3-$\sigma$ uncertainties of the fitted transit times. 

\subsection{Stacked transit analysis \label{sec:stacked_analysis}}An independent transit analysis of K2-146 b and c was performed using stacked transit light curves. The light curves were cut such that only data within 8 transit durations, centred on each transits were used in our analysis.
The selected transits of K2-146 b and K2-146 c used in our analysis are indicated by red and blue lines, respectively, in Figure \ref{fig:K2_lightcurve}.

We employed a Markov-chain Monte Carlo (MCMC) approach to derive the system parameters of K2-146. The stacked transits of K2-146 b and K2-146 c were modelled using the analytical functions by \cite{2002ApJ...580L.171M}. The transit model was implemented using the package \texttt{PyTransit} \citep{2015MNRAS.450.3233P}, and a quadratic stellar limb-darkening law was applied. The fitted transit parameters are the planet-to-star radius ratios, $k_b$ and $k_c$, the orbital inclinations, $i_b$ and $i_c$, stellar density, $\rho_s$, and the triangle sampling \citep{kipping_2010c} of the quadratic stellar limb-darkening coefficients $u_1$ and $u_2$, where uninformative priors was used. The orbital periods or the planets are kept fixed.

The MCMC method was implemented using the Python package \texttt{emcee} \citep{2013PASP..125..306F} for Bayesian parameter estimation. A $\chi^2$ statistics was used for likelihood estimation in our model. We computed the log-likelihood, $\log L$, following Equation \ref{eq:Log}. An initial burn-in phase of 20 MCMC chains $\times$ 10000 steps was implemented to optimize the convergence of the fit. To obtain reasonable uncertainties in the transit parameters, we rescaled the error bars such that the value of the reduced $\chi^2$ equals to 1. We then initiated 100 MCMC chains of $5\times10^4$ steps to sample the posterior space. We checked for convergence and discarded the first 2000 steps, then adopted the median, $16^{\rm th}$, and $84^{\rm th}$ percentiles of the samples in the marginalized posterior distributions as the fitted values and their 1-$\sigma$ uncertainties. The results of the fitted transit parameters of K2-146 b and K2-146 c are presented in appendix Table \ref{tab:transitparams}. The best-fitted transit parameters of K2-146 b and K2-146 c obtained here are generally consistent within $\sim$1-$\sigma$ with those derived by \texttt{PyTV} .

\section{TTV Analysis and results \label{sec:TTV_analysis}}

We modeled the TTVs for the two planets using the {\tt TTVFast} code \citep{2014ApJ...787..132D}, considering Newtonian gravitational interactions between the host star and the two planets alone. For each planet, we fitted planet-to-star mass ratio, orbital period $P$, eccentricity and argument of periastron parameterization ($\sqrt{e}\cos\omega$ and $\sqrt{e}\sin\omega$, so that the uniform priors on these parameters correspond to the priors flat in $e$ and $\omega$), and time $t_0$ of inferior conjunction closest to the dynamical epoch $t_{\rm epoch}(\mathrm{BJD}-2454833)=3467.8$. The elements are osculating Jacobi elements defined at $t_{\rm epoch}$, and the time $t_0$ is related to the time of periastron passage $\tau$ via 
$2\pi (t_0-\tau)/P = E_0 - e\sin E_0$, where $E_0=2\arctan\left[\sqrt{{1-e}\over{1+e}}\tan\left({\pi \over 4}-{\omega \over 2}\right)\right]$. Considering that both planets are transiting in the C18 data, which are close to the dynamical epoch, the inclination and longitude of ascending node at the epoch were fixed to be $\pi/2$ and $0$, respectively. The likelihood was defined using the usual $\chi^2$ as $\exp(-\chi^2/2)$, where we directly adopted the errors from {\tt PyTV} because the scatter in the data around the best model was found to be consistent with the assigned values. 
We adopted uniform priors for all these parameters and used {\tt emcee} \citep{2013PASP..125..306F} to sample from their posterior distribution. 

\begin{figure}[htbp!]
\gridline{\fig{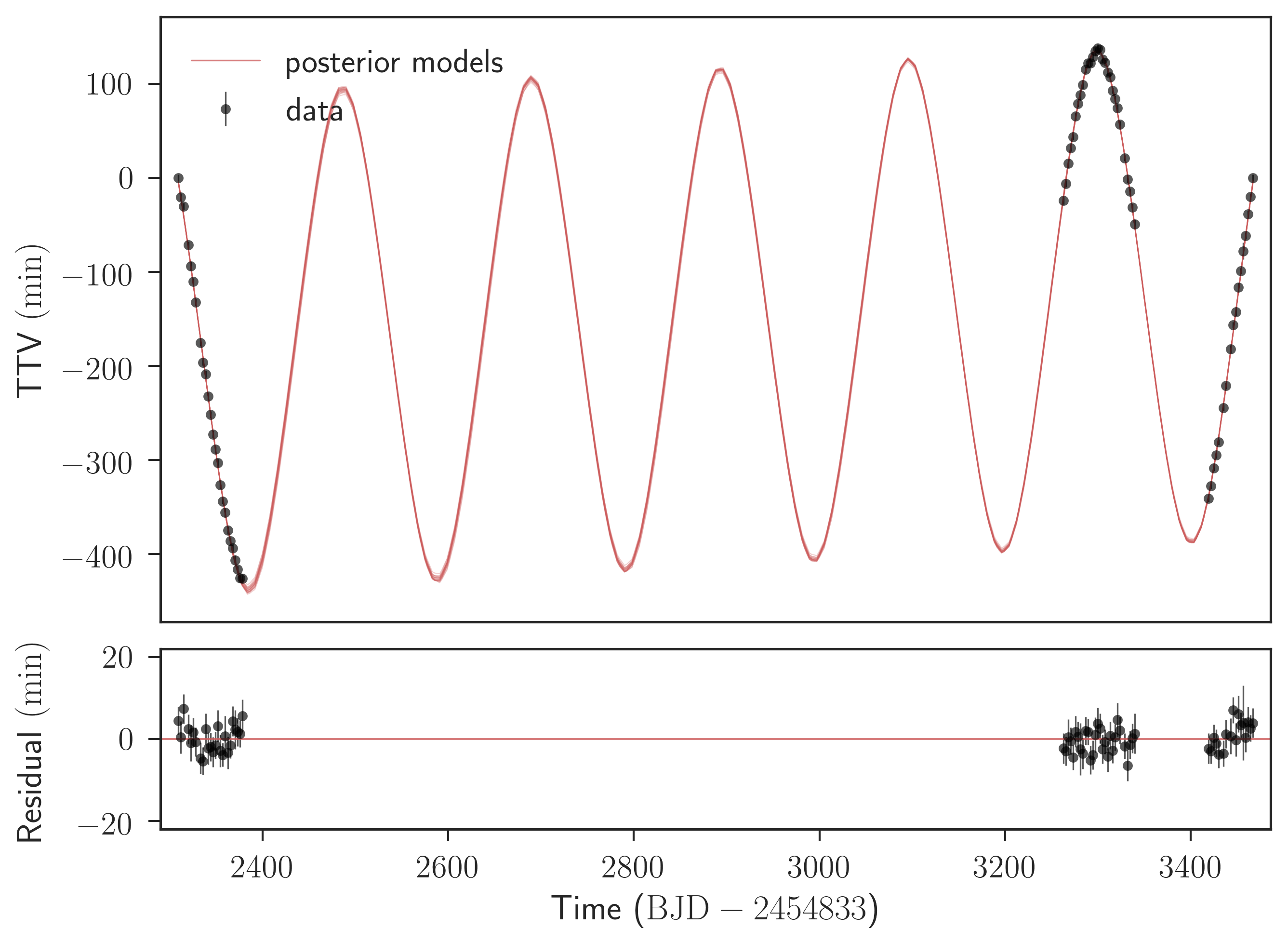}{0.45\textwidth}{(a) K2-146 b}}
\gridline{\fig{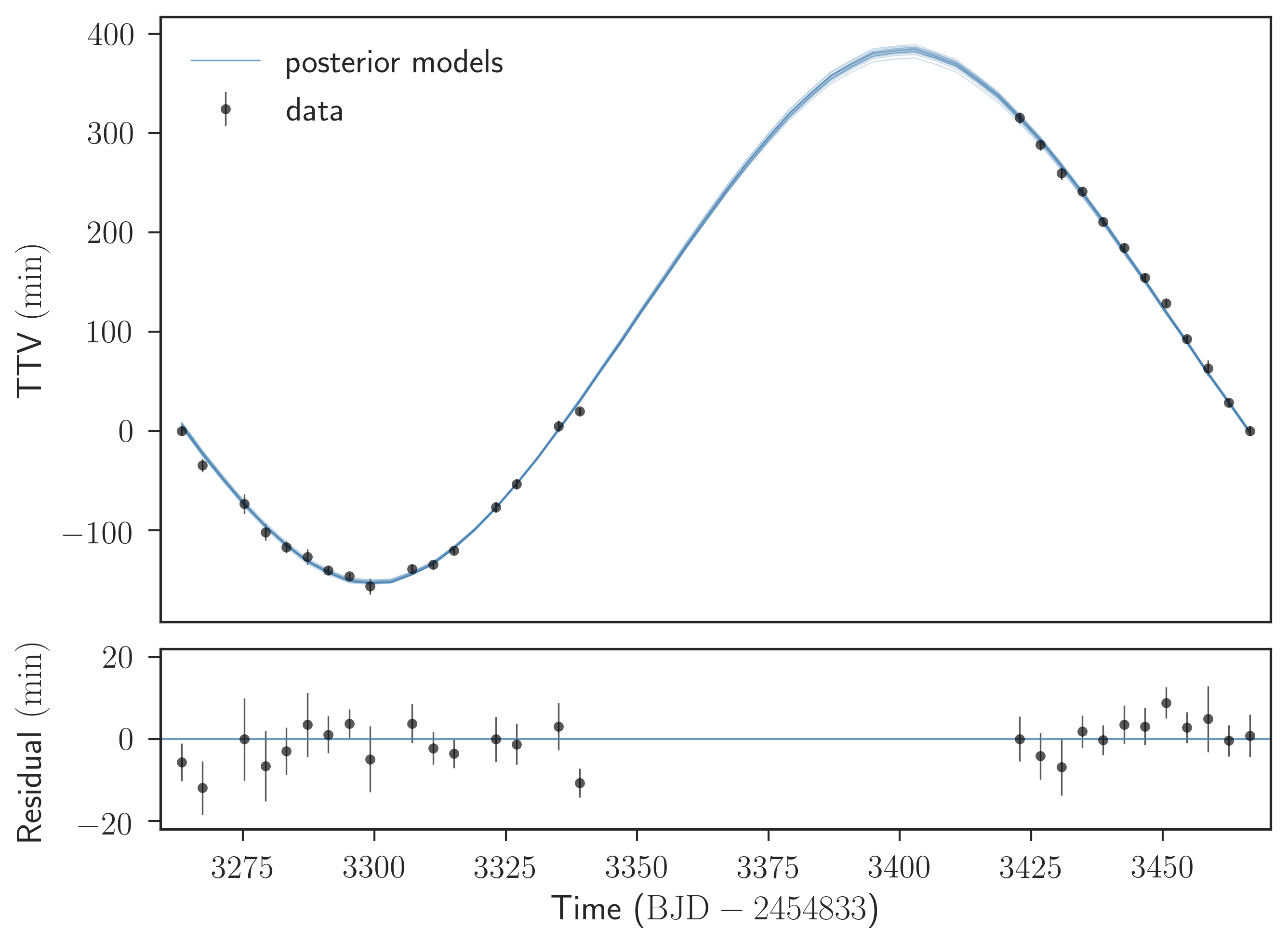}{0.45\textwidth}{(b) K2-146 c}}
\caption{Observed and modeled TTVs of K2-146 b (top) and c (bottom). Here TTVs are plotted with respect to the epoch $\mathrm{BJD}=2454833+3467.4374$ and mean period $2.65696\,\mathrm{days}$ for planet b; the epoch $\mathrm{BJD}=2454833+3466.6029$ and mean period $3.98579\,\mathrm{days}$ for planet c.
Thin blue lines are 20 random posterior models. Residuals are computed for the best-fit model.}
\label{fig:ttv_models}
\end{figure}

Figure~\ref{fig:ttv_models} shows the TTV models generated with 20 sets of parameters randomly drawn from the posterior distribution. Table~\ref{tab:ttv_params} summarizes the median and 68\% credible interval of the marginal posterior distribution for each parameter: the upper parts show the fitted parameters, and the lower part shows the derived parameters. The planet-to-star mass ratios combined with the host star mass yield planetary masses of $5.6\pm0.7\,M_\oplus$ for K2-146 b and $7.1\pm0.9\,M_\oplus$ for K2-146 c. Moderate eccentricities are favored for both planets, but they show a strong negative correlation and one of the planets may have a nearly circular orbit. The $99.7\%$ upper limit for the eccentricity is $0.3$ for both planets. The observed difference in the arguments of periastron is consistent with anti-alignment of the apses. The implications of these features will be discussed in Section \ref{sec:discussion}.

\begin{deluxetable}{lcc}[!htbp]
\tablecaption{Masses and orbital elements for K2-146 b and c determined from TTV modeling.\label{tab:ttv_params}}
\tablehead{
\colhead{} & \colhead{K2-146 b} & \colhead{K2-146 c}
}
\startdata
{\it Fitted parameters}\\
\ \ $M_{\rm p}/M_{\rm s}$ [$\times10^{-5}$] & $4.7\pm0.2$ & $6.0\pm0.2$\\
\ \ $P$ [days] & $2.6698\pm0.0001$  & $3.9663\pm0.0002$\\
\ \ $\sqrt{e} \cos \omega$  &  $-0.36^{+0.11}_{-0.08}$    & $0.40^{+0.08}_{-0.10}$ \\
\ \ $\sqrt{e} \sin \omega$ & $-0.07^{+0.09}_{-0.08}$       & $0.01\pm0.07$\\
\ \ $t_{0}$ [$\mathrm{BJD}-2454833$] & $3467.4345\pm0.0007$ &  $3466.6019\pm0.0009$ \\
{\it Derived parameters}\\
\ \ $M_{\rm p}$ [$M_\oplus$]\tablenotemark{a}
& $5.6\pm0.7$ & $7.1\pm0.9$\\
\ \ $e$ & $0.14\pm0.07$    & $0.16\pm0.07$\\
\ \ $\omega$ [deg] & $191^{+11}_{-15}$ & $2^{+12}_{-10}$\\
\enddata
\tablenotetext{a}{Derived using $M_{\rm s}=0.358\pm0.042\,M_\odot$.}
\tablecomments{The values quoted here are the medians and symmetric 68\% credible intervals of the marginal posteriors. The orbital elements are defined at the epoch $t_{\rm epoch}(\mathrm{BJD}-2454833)=3467.8$.}
\end{deluxetable}

\section{Transit modelling} \label{sec:tlcm_analysis}
\subsection{Updating transit parameters}
The stacked transit light curves of K2-146 b and c were re-analyzed incorporating the eccentricity information from the TTV analysis. The transits were modelled with \texttt{TLCM} (Transit and Light Curve Modeller; Csizmadia 2019, MNRAS, \textit{under review}), a software tool for joint radial velocity and transit light curve fit, or transit light curve fit only. It utilizes the \cite{2002ApJ...580L.171M} and \cite{2013PASP..125...83E} subroutines to calculate the transit light curve shapes for every time moment when we have an observation. 
A wavelet-filter \citep{2009ApJ...704...51C} can be applied to model the red-noise effects. 
Contamination is also taken into account. The light curve part uses a quadratic limb darkening law. The stellar radius, based on the value measured by \cite{2018AJ....155..127H} as well as the spectroscopic $\log g$ values can be used as priors for the fit.
Parameter estimation is done via Genetic Algorithm, refined by Simulated Annealing \citep{Geem2001}.
The final parameter estimation is done by using several chains of MCMC with at least $10^5$ steps. The median and the width of the chains will define the finally adopted solutions and its uncertainty ranges. The Gelman-Rubin statistic (e.g. \cite{2006PASP..118.1351C}) is used to check the convergence of chains. For detailed descriptions of TLCM, we refer the reader to the following works: \cite{2011A&A...531A..41C,2015A&A...584A..13C,2017MNRAS.464.2708S}. 

\begin{figure}[htbp!]
\gridline{\fig{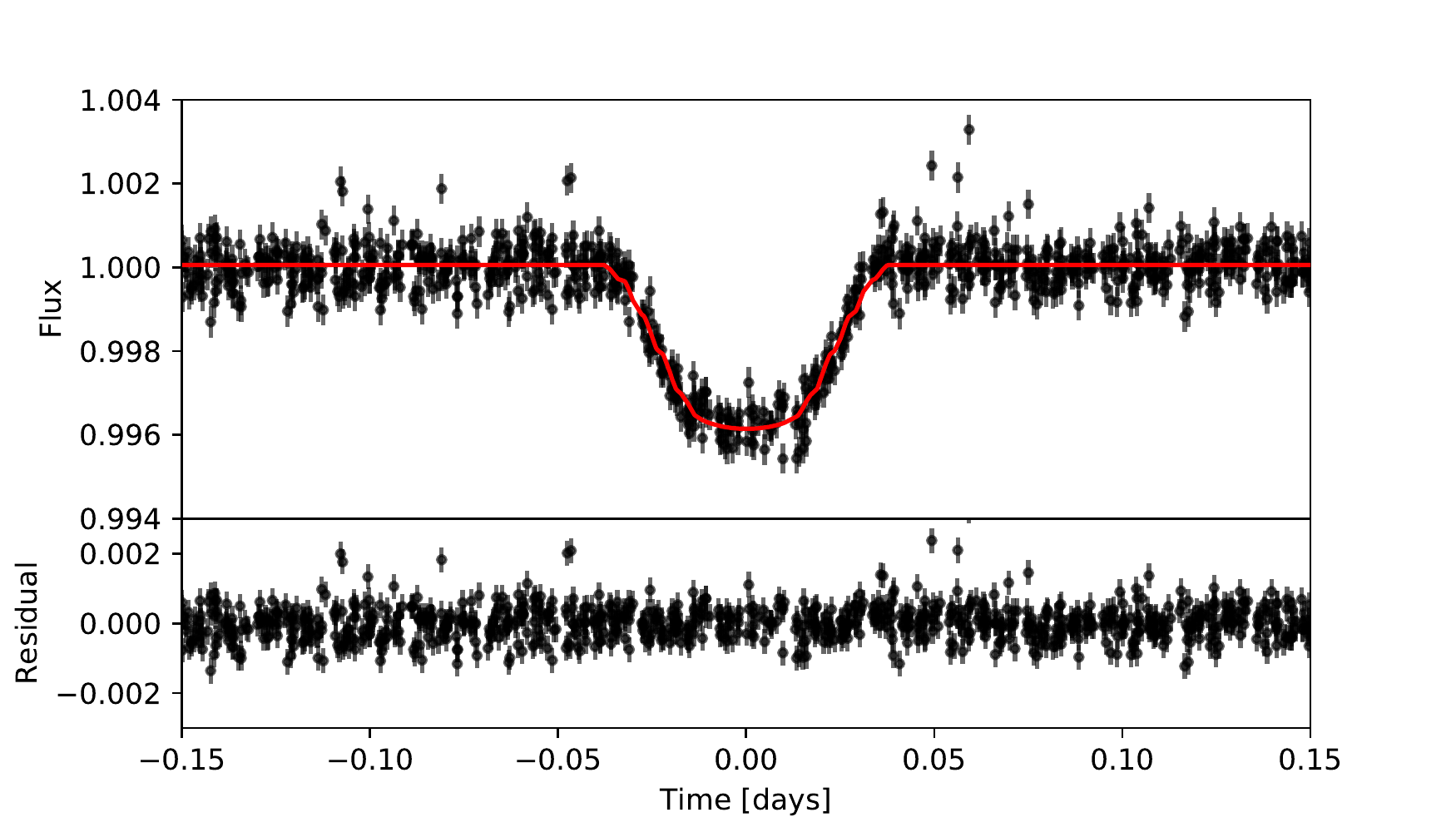}{0.5\textwidth}{(a) K2-146 b}}
\gridline{\fig{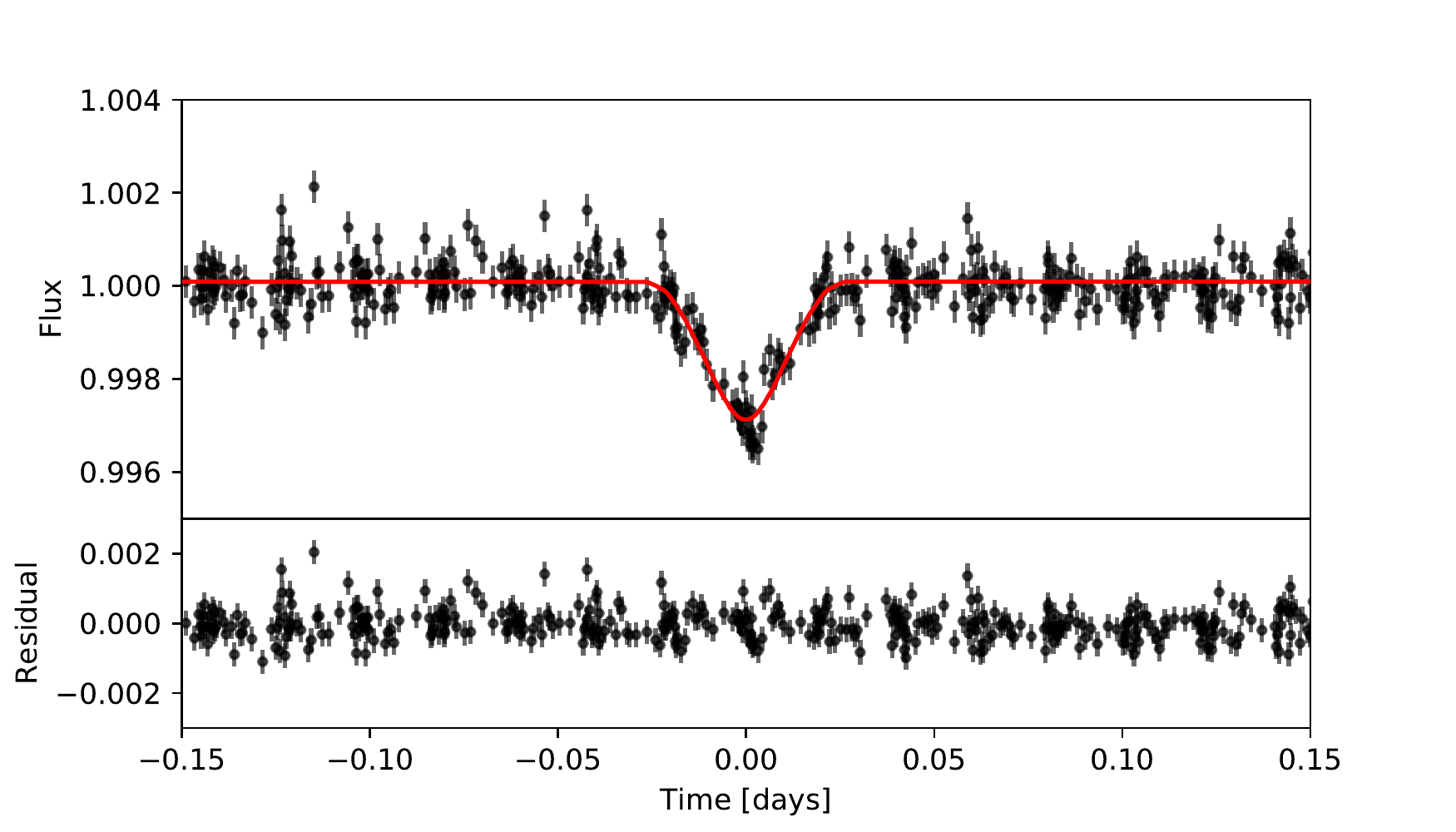}{0.5\textwidth}{(b) K2-146 c}}
\caption{Stacked light curves showing the TTV-corrected transits of (a) K2-146 b, and (b) K2-146 c in the top panels. The red lines are the best-fit transit model and the corresponding residuals are shown in the bottom panels. }
\label{fig:tlcm_lightcurve}
\end{figure}

The fitted parameters are the epoch of mid-transit, $T_0$, the scaled semi-major axis ($a/R_s$), planet-to-stellar radius ratio ($R_p/R_s$), the impact parameter, $b$, and the quadratic stellar limb-darkening coefficients $u_{+} = u_1 + u_2$, and $u_{-} = u_1 - u_2$. The orbital periods, $P$, of the planets were kept fixed. In addition, we used the eccentricity and argument of periastron derived from Section \ref{sec:TTV_analysis} as priors to perform our analysis. Table \ref{tab:tlcm_params} presents the best-fit transit parameters of K2-146 b and K2-146 c. The resulting best-fit transit model of the inner and outer planets are shown in Figure \ref{fig:tlcm_lightcurve}. 

\begin{table*}[htbp!]
\centering
\caption{Best-fit planet parameters of K2-146 b,c from a stacked transit analysis, and their corresponding $1-\sigma$ uncertainties. The orbital periods, $P$ of the planets were kept fixed at the values derived from the TTV analysis.\label{tab:tlcm_params}}
\begin{tabular}{llcc}
\hline
\hline          
Parameter                                   & Description [unit]                         & \multicolumn{2}{c}{Values and uncertainties} \\
\hline          
                                            &                                     & K2-146 b              & K2-146 c             \\
\hline          
$P$   & Period [day]                    &   $2.6698$                & $3.9663$  \\
$T_0$ & Epoch [day from transit center] &   $0.00009 \pm 0.00038$   & $-0.00015 \pm 0.00076$ \\
$R_p$ & Radius [\Rearth]                &   $2.25 \pm 0.10$         & $2.59^{+1.81}_{-0.39}$ \\
$a$   & Semi-major axis [AU]            & $0.0248 \pm 0.0002$       &  $0.0327 \pm 0.0006$ \\
$b$   & Impact parameter                &  $0.391 \pm 0.069$       &  $0.930 \pm 0.097$ \\
$i$   & Inclination [\degree]           &  $88.5 \pm 0.3$           &  $87.3 \pm 0.3$  \\
$R_p/R_s$   & Scaled planet radius              & $0.0589 \pm 0.0014$ & $0.0680^{+0.1226}_{-0.0254}$ \\
$a/R_s$   & Scaled semi-major axis              & $15.250 \pm 0.126$ & $20.064 \pm 0.412$ \\
$u_+ = u_1+u_2$  & Combined limb-darkening coefficient & $0.575 \pm 0.171$ & $0.678 \pm 0.169$ \\
$u_- =u_1-u_2$   & Combined limb-darkening coefficient &  $0.022 \pm 0197$ & $0.072 \pm 0.210$ \\
$\rho_p$  & Density [\gcm]   &  $2.702 \pm 0.494$  &   $2.246^{+1.883}_{-1.846}$  \\ 
\hline          \\
\end{tabular}
\end{table*}

\subsection{Planet impact parameters}
The outer planet K2-146 c was only found to transit in the latter campaigns C16 and C18. This suggests orbital plane precession is at play. We searched for seasonal changes in the impact parameters of both planets. The transits of the inner and outer planets were stacked separately in each \texttt{K2} campaign. The mean orbital elements of each campaign were obtained from the posterior samples of the TTV analysis. We fixed $\sqrt{e} \sin \omega$ and $\sqrt{e} \cos \omega$ values on these mean values for the analysis of separate campaigns. We then ran TLCM on the campaign-stacked data, the free parameters in each campaign were the scaled semi-major axis, impact parameter, radius ratio, and epoch. The radius ratio was the same from campaign to campaign. 

For planet b, we obtained an impact parameter of $b = 0.42\pm0.11$, $b = 0.25\pm0.10$, and $b = 0.41\pm0.15$, for C05, C16, and C18 respectively. While the impact parameter is practically the same in C05 and C18, it differs by approximately 1-$\sigma$ in C16 in comparison to the two other campaigns. This indicates that we do not see a significant change in the impact parameter of planet b in these data. In the case of planet c, we obtained an impact parameter of $b = 0.94\pm0.10$ and $b = 0.92\pm0.10$ for C16 and C18, respectively. Again we do not see significant changes in the impact parameter of the planet, although the null detection of transits in C05 does suggest that it has been drifting. We attribute this to a shorter time baseline between C16 and C18 ($\sim200\,\mathrm{days}$) compared to $\sim1000\,\mathrm{days}$ separation between C05 and C16.

\section{Discussion \label{sec:discussion}} 

\subsection{Dynamic stability of the K2-146 system} 
We investigated the dynamical stability for the K2-146 multi-planet system, independent of the TTV analysis, to obtain the mass limits on the two Neptune-size planets and to establish the stability of the system.

A Hill-sphere is the region where the planet's gravity is dominating over the central star. If the Hill-spheres overlap then there is a chance to collide or to remove one of the planets from the system. Using the orbital parameters reported in Table \ref{tab:ttv_params}, we find that planets $b$ and $c$ have Hill-sphere radii of 2.1\% and 2.2\% of their semi-major axes which suggests that the planets are likely stable.

The N-body simulation code \texttt{Mercury6} \citep{1999MNRAS.304..793C} was used to study the orbital evolution and determine the dynamic stability of the system. We chose the `hybrid symplectic and Bulirsch-Stoer integrator' mode of \texttt{Mercury6} to compute close encounters in the system. We adopted the stellar mass and radius reported in \cite{2018AJ....155..127H} for the central star.

We employed 5105 integrations, each with an integration period of 2 Myr. An initial step size of 2.7 d was selected, subsequent step sizes were adjusted by the variable time-step algorithms in the program to maintain integration accuracy. The orbital parameters of the system were recorded every 2 years. For each integration, the orbital periods ($P_b$ and $P_c$) were chosen from a Gaussian-distribution with the center and 1$-\sigma$ reported by \texttt{PyTV} in Section \ref{sec:timing_extraction}. The eccentricities of the planets ($e_b$ and $e_c$) were drawn from a uniform distribution between $e=0$ and $e=0.3$.
The two planets were assumed to be co-planar with fixed inclinations and longitudes of the ascending node where $i_b,i_c = 90^{\circ}$ and $\Omega_b, \Omega_c = 0^{\circ}$. 
The planetary masses of K2-146 b and K2-146 c ($M_b$ and $M_c$ respectively) were at first randomly drawn from a half-normal distribution with a width of $1000$ \Mearth. We further increased the number of data points in the lower mass regions by uniformly sampling the 0-30 \Mearth\ range (the 0-3 \Mearth\ was even more densely sampled). All samples were merged in the presented final result. 
The arguments of periastron ($\omega_b$ and $\omega_c$) and mean anomalies of the planets were drawn from a uniform sample where $0^{\circ} < \omega_b, \omega_c < 360^{\circ}$ and $0^{\circ} < \mathcal{M}_b, \mathcal{M}_c < 360^{\circ}$. In each simulation, the system becomes unstable when (1) the planets collides with one another or with the central star; or (2) the planets are ejected from the system, i.e. $e_b,e_c > 1$ and/or $a > 10$ au.

\begin{figure}[htbp!]
\begin{center}
\includegraphics[width=0.45\textwidth]{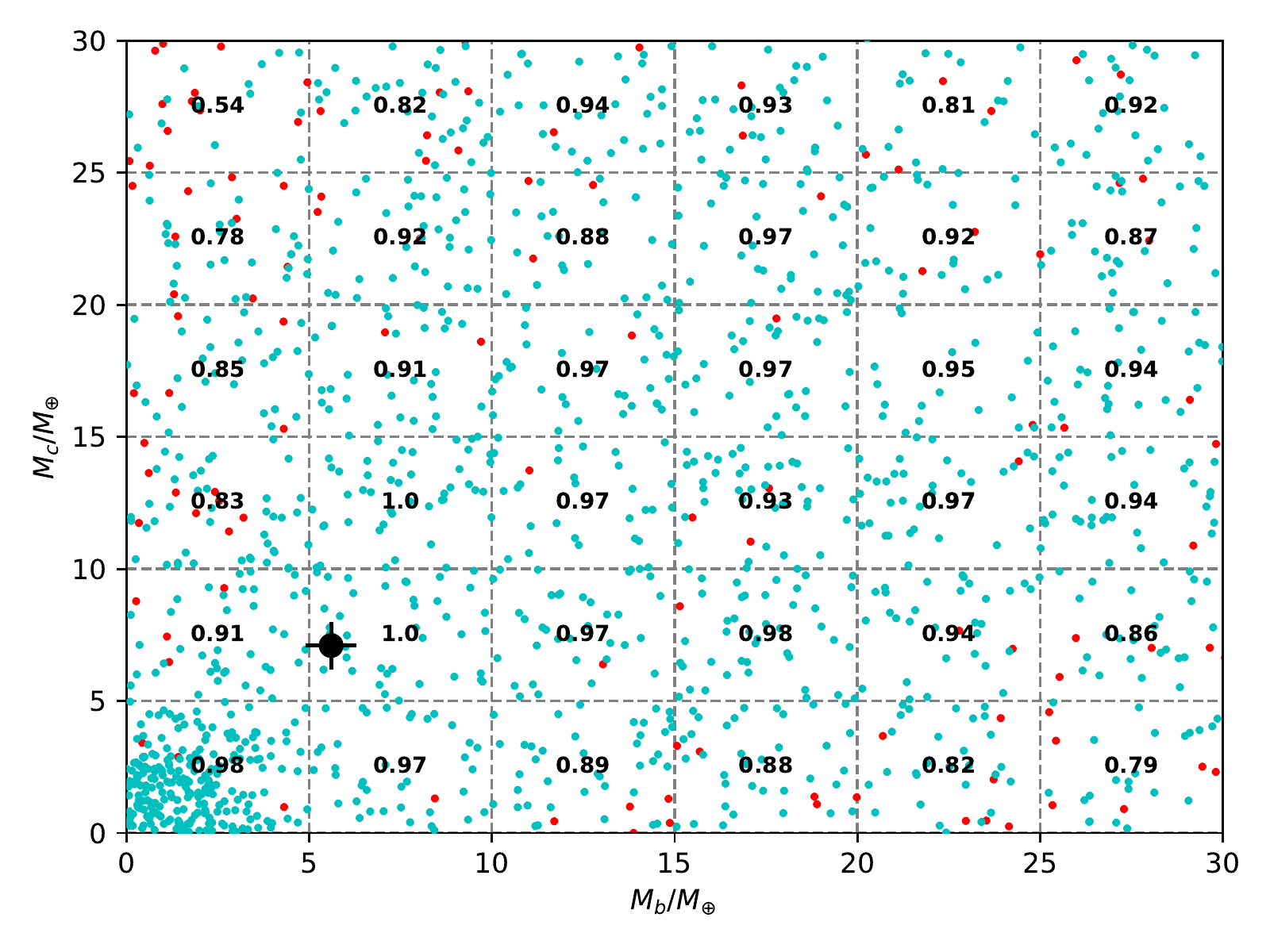}
\caption{Mass-Mass plot of all dynamical simulation of the multi-planet system K2-146 for 2 Myr centred on the sample space where $\rm M_b,M_c < 30$ \Mearth. The mass distribution of K2-146 b and K2-146 c are shown along the x- and y-axis, respectively. The cyan and red circles represent systems which are dynamically stable and unstable, respectively. The black circle denotes masses of planet b and c derived in our TTV analysis. The mass-mass plot is divided into grids with widths of 5 \Mearth. The fraction of stable orbital configurations of each grid is calculated and overplotted. Over 90\% of the systems can remain dynamically stable for 2 Myr if the mass ratio of the two planets is close to unity. \label{fig:stability_plot}}
\end{center}
\end{figure}

Our results presented in Figure \ref{fig:stability_plot} shows that a significant fraction of systems are dynamically stable for 2 Myr for planets with masses up to $\sim$30 \Mearth. The stable and unstable configurations are denoted by cyan and red circles, respectively. We calculated the fraction of stable orbits within each grid with widths of 5 \Mearth. We found that over 90\% of the systems remained stable for 2 Myr if the mass ratio of K2-146 b and K2-146 c is close to unity.
The upper mass limits of the planets obtained from dynamical constraint is consistent with the planet masses derived from TTVs in Section \ref{sec:TTV_analysis}. The mass ratio of the planets determined from the TTV analysis is $0.783\pm0.006$. The system is thus very likely to be stable on a timescale of at least 2 Myr.

\subsection{Orbital resonance of the sub-Neptunes} 

The orbital periods of K2-146 b and K2-146 c show a 3:2 commensurability. To assess whether the planet pair is truly trapped in a 3:2 MMR, we monitored the orbital evolution simulation of the so-called resonance angles over a 2000 years long interval.

The resonance arguments, $\Theta_1$ and $\Theta_2$, for a pair of planets in a 3:2 MMR are defined as
\begin{equation}
\begin{aligned}
\Theta_1 &= (p+q) \cdot \lambda_c - p \cdot \lambda_b  - q \cdot \varpi_b, \\
\Theta_2 &= (p+q) \cdot \lambda_c - p \cdot \lambda_b  - q \cdot \varpi_c, \\
\end{aligned}
\end{equation}
so the difference is only in the last term. The mean motion resonance is defined as
\begin{equation}
    \frac{n_c}{n_b} = \frac{P_b}{P_c} = \frac{p}{p+q} 
\end{equation}
therefore in our case $p=2$ and $q=1$.
The quantity $\lambda = \mathcal{M} + \varpi$ is often called the mean longitude, and $\mathcal{M}$ is the mean anomaly. The longitude of pericenter is $\varpi = \omega + \Omega$ (e.g. \citealt{1999ssd..book.....M}). The resonant angles measure the angle between the two planets at the conjunction point. If any resonant angle librates rather than circulates, then the planets are in mean motion resonance. 

We performed twenty thousand simulations to evaluate the resonance angles of the planet pair. In each simulation, the planet to star mass ratios and orbital elements of the planets were drawn from the posterior samples of the TTV analysis. Stellar mass in Table \ref{tab:stellarparam} was used to convert the mass ratios into planetary masses. Then the orbits were numerically integrated for 2000 years. The values of the relevant parameters were saved for every 36 days of the integration, and then the resonant angles were calculated. For each integration, we recorded the maximum amplitude of the resonant angles. We took the modulo $360$\degree\ values of the resonant variables. 

\begin{figure}[htbp!]
\gridline{\fig{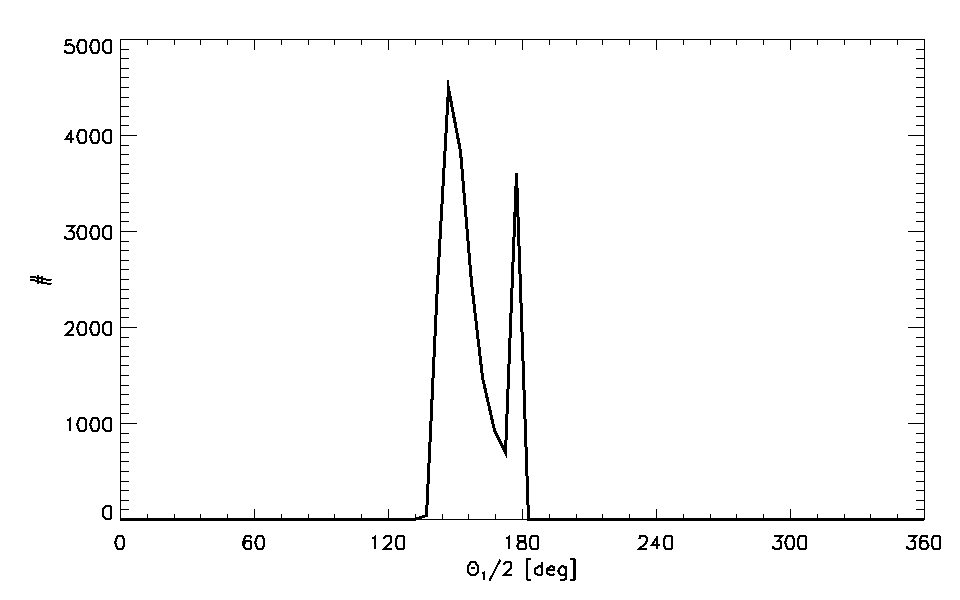}{0.45\textwidth}{(a) Resonance angle $\Theta_1 = 3\lambda_c - 2\lambda_b - \varpi_b$}}
\gridline{\fig{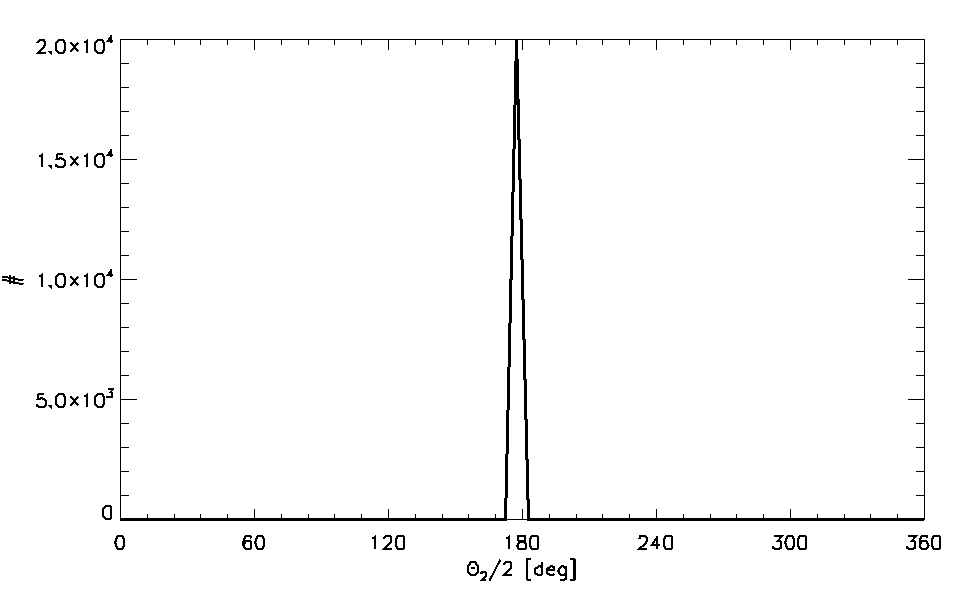}{0.45\textwidth}{(b) Resonance angle $\Theta_2 = 3\lambda_c - 2\lambda_b - \varpi_c$}}
\caption{(a) The histograms of the half-maximum amplitude of the resonant angles $\Theta_1$ (Top) and $\Theta_2$. 77\% of the 20000 simulations with orbital elements drawn from the TTV posterior samples have $\Theta_1 \sim 150$ \degree, corresponding to orbits librating in a horseshoe orbit. \label{fig:resonance_histogram_plots}}
\end{figure}

\begin{figure}[htbp!]
\begin{center}
\includegraphics[trim=20 0 0 30,clip,width=0.45\textwidth]{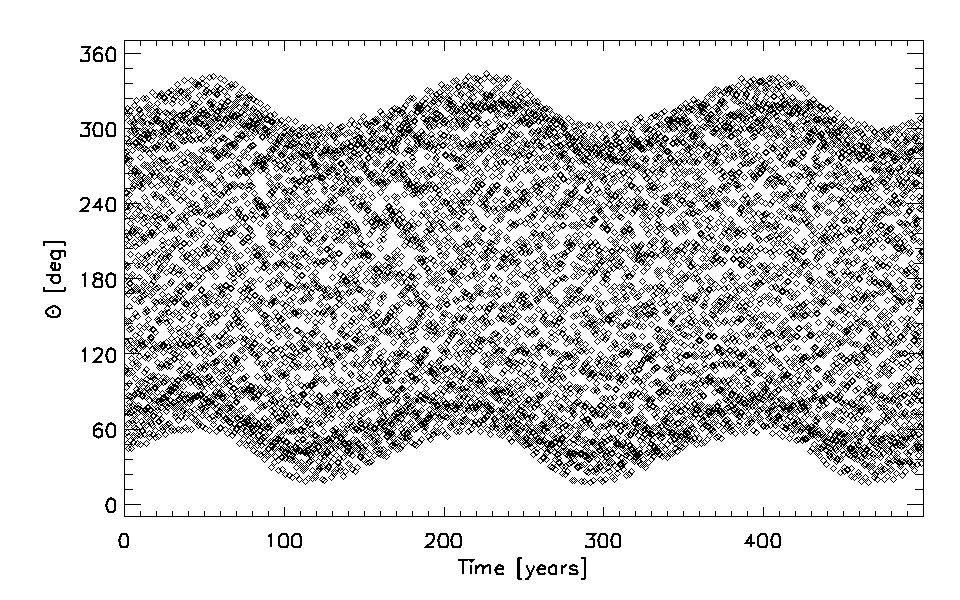}
\caption{Evolution of the orbital solution giving the smallest libration half-amplitude. This variability of $\Theta_1$ means that the planet $c$ is ahead or behind the conjunction point by a maximum of $\sim$150\degree, so it librates around the conjunction point. \label{fig:resonance_plot}}
\end{center}
\end{figure}

Figure \ref{fig:resonance_histogram_plots} shows the histograms of the maximum half-amplitude of the resonant angles. The largest peak at around $\Theta_1/2=150$\degree\ is due to libration, so planet c is librating with a half-amplitude of $\sim150$\degree\ around the conjunction points at the time of predicted conjunction times (every third orbital cycle). The smaller, narrower peak at $\Theta_1/2=180$\degree\ corresponds to either horseshoe-shaped orbits or $\Theta_1$ circulates and then $\Theta_2$ librates. We found approximately 77\% of simulations show a libration amplitude of around 150\degree\, which implies there is at least a 77\% chance that the system is trapped in 3:2 MMR instead of a random commensurability. Figure \ref{fig:resonance_plot} shows the simulation which gave the smallest libration angle. 

The width of the peak in these histograms are caused by the uncertainties of the masses and orbital elements derived from TTVs. To increase their precisions, more transit observations are needed from this system. Despite the faintness of the host star, which leads to smaller accuracy in TTVs, it would be worthy to try to observe it with high cadence with \textit{TESS} \citep[Transiting Exoplanet Survey Satellite;][]{2014SPIE.9143E..20R} and \textit{PLATO} \citep[PLAnetary Transits and Oscillations of stars;][]{2014ExA....38..249R}. Alternatively, observations may also be obtained with the less precise ($\sim$10 minutes) timing values from \textit{CHEOPS} \citep[CHaracterising ExOPlanet Satellite;][]{2013EPJWC..4703005B}, because the precision could be compensated with many hours of TTV-values in this case.

\subsection{Planet composition}
Our light curve analysis in section \ref{sec:tlcm_analysis} and TTV analysis in section \ref{sec:TTV_analysis} reveal that K2-146 b has a mass and radius of $2.25 \pm 0.10$ \Rearth\ and $5.6 \pm 0.7$ \Mearth, respectively, corresponding to a bulk density of $2.702 \pm 0.494$ \gcm. The mass and radius of the outer planet K2-146 c are $2.59_{-0.39}^{+1.81}$ \Rearth\ and $7.1 \pm 0.9$ \Mearth, respectively, which correspond to a bulk density of $2.246_{-1.846}^{+1.883}$ \gcm. K2-146 b and K2 -146 c are orbiting at a distance of 0.0248 AU and 0.0327 AU, respectively. We assumed the planets have an albedo of 0, and a re-radiation factor of 1/4, where atmospheric circulation redistributes the energy around the planetary atmosphere, then re-radiate the energy back into space. Under these assumptions, the equilibrium temperatures of planet b and planet c are approximately 590 K and 520 K, respectively.

\begin{figure}[htbp!]
\begin{center}
\includegraphics[width=0.5\textwidth]{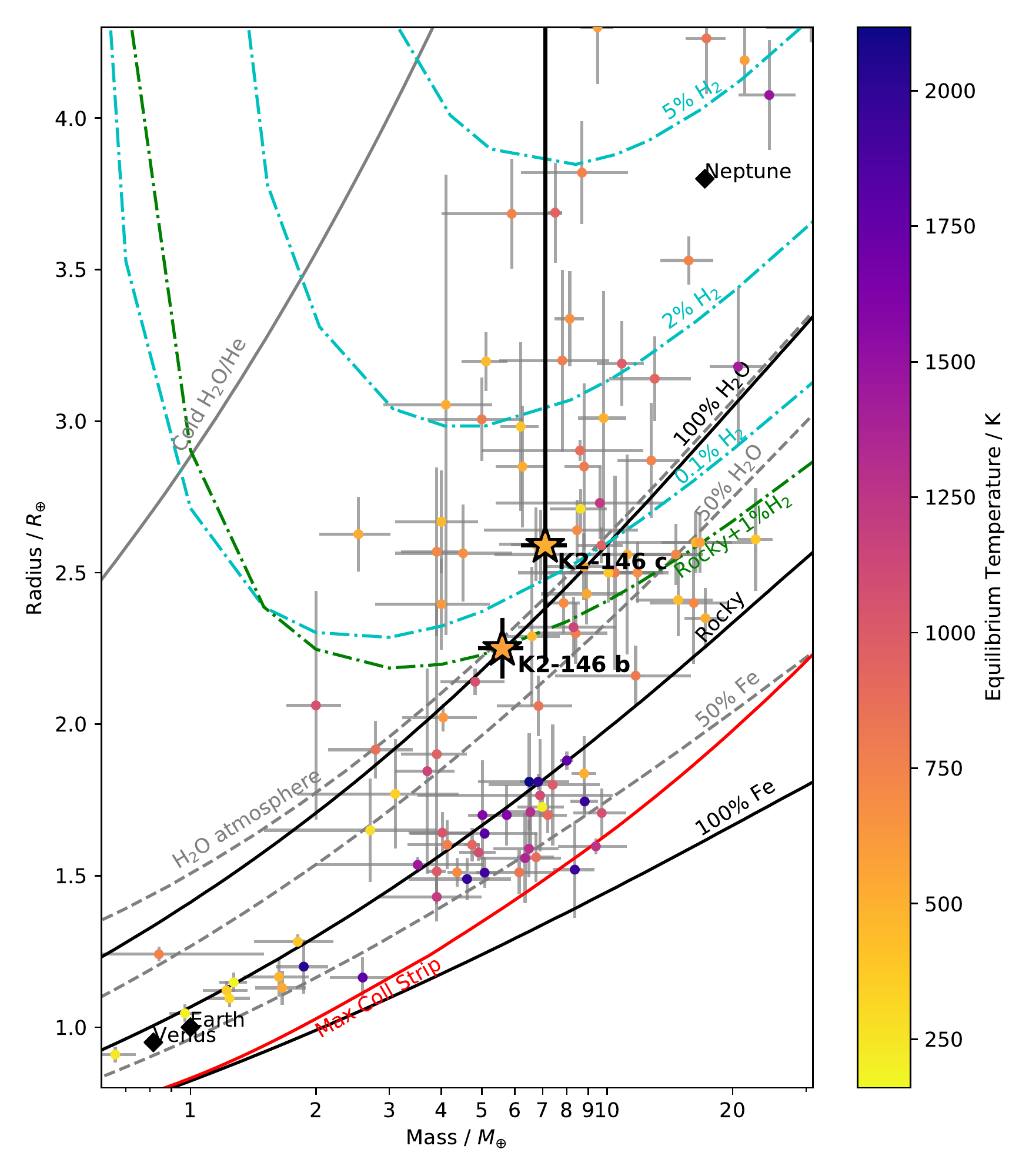}
\caption{Mass-Radius plot of known planets with masses constrained to a precision of better than approximately 30\%. The masses and radii of K2-146 b and c are indicated by the stars. The colors of each data point shows the planet equilibrium temperature as indicated by the color bar on the right. The mass-radius relations of small planets of different compositions are taken from \cite{2016ApJ...819..127Z,2019arXiv190604253Z}. The different compositions are indicated by the solid (100\% water, 100\% rock, or 100\% iron), dashed (mixtures of water, rock and iron) and dashed-dot lines (water-rich cores with a hydrogen envelope or Earth-like rocky cores with a hydrogen envelope). Solar system planets are labeled with black diamonds. The red solid line gives the minimum radii of rocky planets constraint from a giant impact model \citep{2010ApJ...712L..73M}. \label{fig:MR_plot}}
\end{center}
\end{figure}

Figure \ref{fig:MR_plot} shows a mass-radius plot of known planets with $M_p < 30$ \Mearth\ where the mass of the planets are determined with a precision better than approximately 30\%. The solid, dashed and dashed dot lines represents the mass-radius relations for different planetary compositions as derived in \cite{2016ApJ...819..127Z,2019arXiv190604253Z}. The color of each data point indicates a planet's equilibrium temperature corresponding to the colour bar. The masses and radii of K2-146 b and c are consistent with cases of a 100\% H$_2$O interior, a water-rich core with the addition of H$_2$O-gaseous atmosphere, or an Earth-like rocky interior with a small fraction of H$_2$-envelope. The large radius uncertainty of K2-146 c means that it could have a more massive H$_2$-envelope. 

\subsection{Formation and evolution}
Observational evidence of planet orbital architectures allows one to place constraints on the formation and dynamical evolution of the system. In the case of K2-146, we observed a number of interesting traits:

1. \textbf{Masses and radii of the planets are consistent with a water-dominated core and the presence of water or H$_2$ envelope} -- it was previously proposed that \textit{in situ} formation of mini Neptunes and super Earths is possible if 50-100\Mearth\ of rocky material is delivered to the inner disk for planet assembly \citep{2012ApJ...751..158H}. However, the large fraction of solids would drift towards the host star in a relatively short timescale, preventing \textit{in situ} formation. Furthermore, \textit{in situ} formation is unlikely to produce significant fraction of atmospheric masses for close-in planets \citep{2014ApJ...795L..15S,2015MNRAS.448.1751I}. Therefore, close-in planets with an atmosphere were likely formed at larger separation from the star in the presence of a gas disk. Subsequently, planets accrete gaseous envelopes as they migrated inwards towards their current locations.

2. \textbf{Evolution of resonance arguments for the planet pair suggests that K2-146 b and c are likely trapped in 3:2 mean motion resonance} -- during formation, planets interact with protoplanetary disks. This drives the migration of planets inward through the disk due to exchange in angular momentum \citep{1979ApJ...233..857G,1980ApJ...241..425G,1979MNRAS.186..799L}. Convergent migration can occur in one of two ways: (1) planets formed at wide separations can move towards one another with different migration speeds; (2) Planets formed in close proximity are massive enough to form a gap in the disk where inner and outer disks would push the planets towards each other. When the planet orbital periods approaches a commensurability, dynamical interactions that follow cause planets to migrate collectively inwards while preserving period commensurability \citep{2001A&A...374.1092S}. Under favourable disk parameters, planet masses and migration speeds, the planet pair can enter a 3:2 MMR after breaking the 2:1 MMR barrier (e.g. \cite{2000MNRAS.313L..47K,2002MNRAS.333L..26N}), such as the case of HD 45364 \citep{2009A&A...496..521C} and KOI-1599 \citep{2019MNRAS.485.4601P}.

3. \textbf{Our TTV model revealed that both the inner and outer planets have moderate eccentricities of $0.14\lesssim e \lesssim0.16$, and are apsidally anti-aligned, i.e. $\Delta \omega = \omega_b - \omega_c \approx 180$\degree} -- convergent migration could have played a role in the observed eccentricity in K2-146 planet pair. After the planets are captured in resonance, the planet pair migrates inwards while maintaining resonance which leads to orbital eccentricity excitation \citep{2002ApJ...567..596L,2013AJ....145....1B}. While the protoplanetary disk is present, the planets could experience eccentricity damping as they migrate. The moderate eccentricities observed in K2-146 b and c implies that the migration process must be fast enough in order to minimize the damping efficiency. The misaligned apsides could have been the result of such migration process. The conjunctions occur when K2-146 c is near periapse and K2-146 b is near apoapse. The longitudes of periapse of both planets are required to precess at the same rate for a stable configuration, such that the lines of apsides is locked in an anti-aligned state \citep{2002ApJ...567..596L}. Due to close proximity of the two planets, such mechanism is present to avoid close encounters. 

4. \textbf{The outer planet K2-146 c showed a change in impact parameter} -- Gravitational interaction between planets can gives rise to apsidal and nodal precession around the host star. The change in impact parameter of the outer planet observed from \texttt{K2} Campaign 5 to Campaign 16 is likely an indication of orbital precession. A misalignment between the planet orbits can be suspected, resulting in a precession of the line of nodes of planet c \citep{2002ApJ...564.1019M}. The precession would then lead to a change in the length of the transit chord. In our TTV analysis, we assumed the planets have coplanar orbits because, if the mutual inclination is large, the two planets are unlikely to transit simultaneously even if their orbits are precessing \citep[cf.][]{2017AJ....153...45M}. This assumption could be tested directly via a joint modeling of TTVs and TDVs, or a photodynamical model of the light curves, which will enable a measurement of the mutual orbital inclination through the constraint on the nodal precession rate (e.g. Kepler-117 \citep{2015MNRAS.453.2644A}, Kepler-108 \citep{2017AJ....153...45M}).

\section{Summary and conclusion \label{sec:conclusion}} 
The strong TTV detected in the mini-Neptune K2-146 b suggested the presence of an additional body in the system. Further photometric observations from \textit{K2} revealed an additional mini-Neptune K2-146 c orbiting at a 4-day period, forming a 3:2 mean-motion resonance with the inner planet. The long observation baseline allowed precise determination of the planet masses via TTV. This demonstrates the importance of follow-up transit observations for parameter and dynamical constraints of a TTV system. 

N-body simulations of K2-146 performed in this work provided a glimpse into the possible stability and resonance configuration of the planet pair. We found that the planets are probably captured into a 3:2 MMR during migration, and that their current orbital configuration can be dynamically stable for at least 2 Myr. Furthermore, the change in the impact parameter of the outer planet suggests some orbital plane precession, resulting in the displacement of the chord of transit and hence the change in transit depth and duration of K2-146 c. This effect can be further investigated using both TTVs and TDVs to constrain the orbit precession rate, and mutual inclination in the system. Further observations with TESS, and in the future PLATO can also provide a better precision in the transit times measurement. A detailed migration model would be valuable to study different precession rates leading to a stable orbital configuration.

Small planets around M dwarfs are frequently found in multi-planet systems. In fact, occurrence studies suggests that there are typically around 2.5 small planets ($R_p < 4$\Rearth) per M dwarfs with periods shorter than 200 days \citep{2015ApJ...807...45D}. Only a small handful of planets, are known around M dwarfs which have planetary radii and masses below the radius and mass of Neptune (e.g. TRAPPIST-1 system; \citealt{2017Natur.542..456G}, LHS-1140 system; \citealt{2017Natur.544..333D,2019AJ....157...32M}, L 98-59 System; \citealt{2019AJ....158...32K}, Gl 357 system; \citealt{2019arXiv190412818L}). The These are laboratories to test planet formation theories and dynamical evolutions, providing clues to the processes involved in building multi-planet systems containing the smallest possible planets.

The expected RV semi-amplitudes of K2-146 b and K2-146 c are $K_b = 5.1 \rm ~m~s^{-1}$ and $K_c = 5.7 \rm ~m~s^{-1}$. The faintness of the host star (J$=12.18$ mag) means that mass measurement by means of RV follow up is challenging for many currently available instruments. Recent precision RV instruments (e.g. Infrared Doppler instrument (IRD); \cite{2014SPIE.9147E..14K}) on 8m class telescopes and future generations of high-resolution infra-red (IR) spectrographs such as CRIRES+ \citep{Dorn2014} on the Very Large Telescope (VLT) are sensitive to cool, low mass M dwarfs which radiate mostly in the IR. RV measurements of the K2-146 system may be within reach.

\acknowledgments
\section*{Acknowledgement}
We thank Mareike Godolt for helpful discussion on interpretation of planet compositions. 
KWFL, JK, SzCs, ME, SG, APH, MP and HR acknowledge
support by DFG grants PA525/18-1, PA525/19-1, PA525/20-1, HA3279/12-1 and RA714/14-1 within the DFG Schwerpunkt SPP 1992, ``Exploring the Diversity of Extrasolar Planets''.
M.F., I.G., and C.M.P. gratefully acknowledge the support of the Swedish National Space Agency (DNR 163/16 and 174/18). 
This work is partly supported by JSPS KAKENHI Grant Numbers JP18H01265 and JP18H05439, and JST PRESTO Grant Number JPMJPR1775.
O.B. acknowledges support from the UK Science and Technology Facilities Council (STFC) under grants ST/S000488/1 and ST/R004846/1.
SM acknowledges support from the Spanish Ministry under the Ramon y Cajal fellowship number RYC-2015-17697.
PK acknowledges support by GACR 17-01752J.
SA and EK acknowledge support by the Danish Council for Independent
Research, through a DFF Sapere Aude Starting Grantnr. 4181-00487B.  SA,
EK, and ML acknowledge support by the Stellar Astrophysics Centre which
funding is provided by The Danish National Research Foundation (Grant
agreement no.: DNRF106).
This work was supported by the KESPRINT collaboration, an international consortium devoted to the characterization and research of exoplanets discovered with space-based missions. \url{www.kesprint.science}
This work has made use of data from the European Space Agency (ESA) mission {\it Gaia} (\url{https://www.cosmos.esa.int/gaia}), processed by the {\it Gaia} Data Processing and Analysis Consortium (DPAC, \url{https://www.cosmos.esa.int/web/gaia/dpac/consortium}). Funding for the DPAC has been provided by national institutions, in particular the institutions participating in the {\it Gaia} Multilateral Agreement. 
This research has made use of NASA's Astrophysics Data System Bibliographic Services, the ArXiv preprint service hosted by Cornell University.



\bibliographystyle{aasjournal}
\bibliography{k2-146_ttv}

\newpage

\appendix

\section{Supplementary figures}

\begin{figure}[htbp!]
\begin{center}
\includegraphics[width=\textwidth]{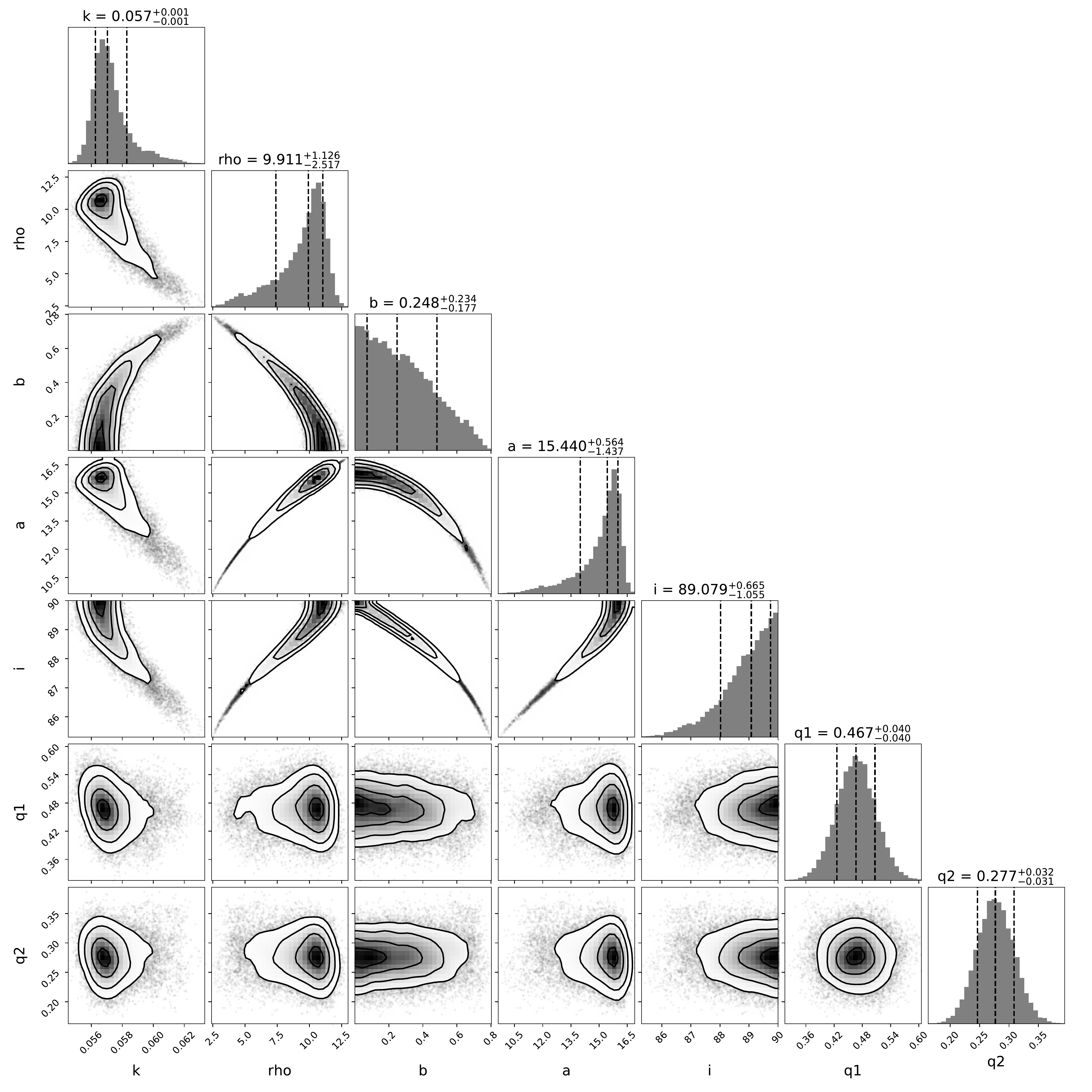}
\caption{Corner plot of the posterior distributions of the transit parameters for K2-146 b. The values above each column are the means of the posterior distributions with their respective 1-$\sigma$ uncertainties. Note that the impact paramter is not well-constrained. \label{fig:lc_cornerplot_b}}
\end{center}
\end{figure}

\begin{figure}[htbp!]
\begin{center}
\includegraphics[width=\textwidth]{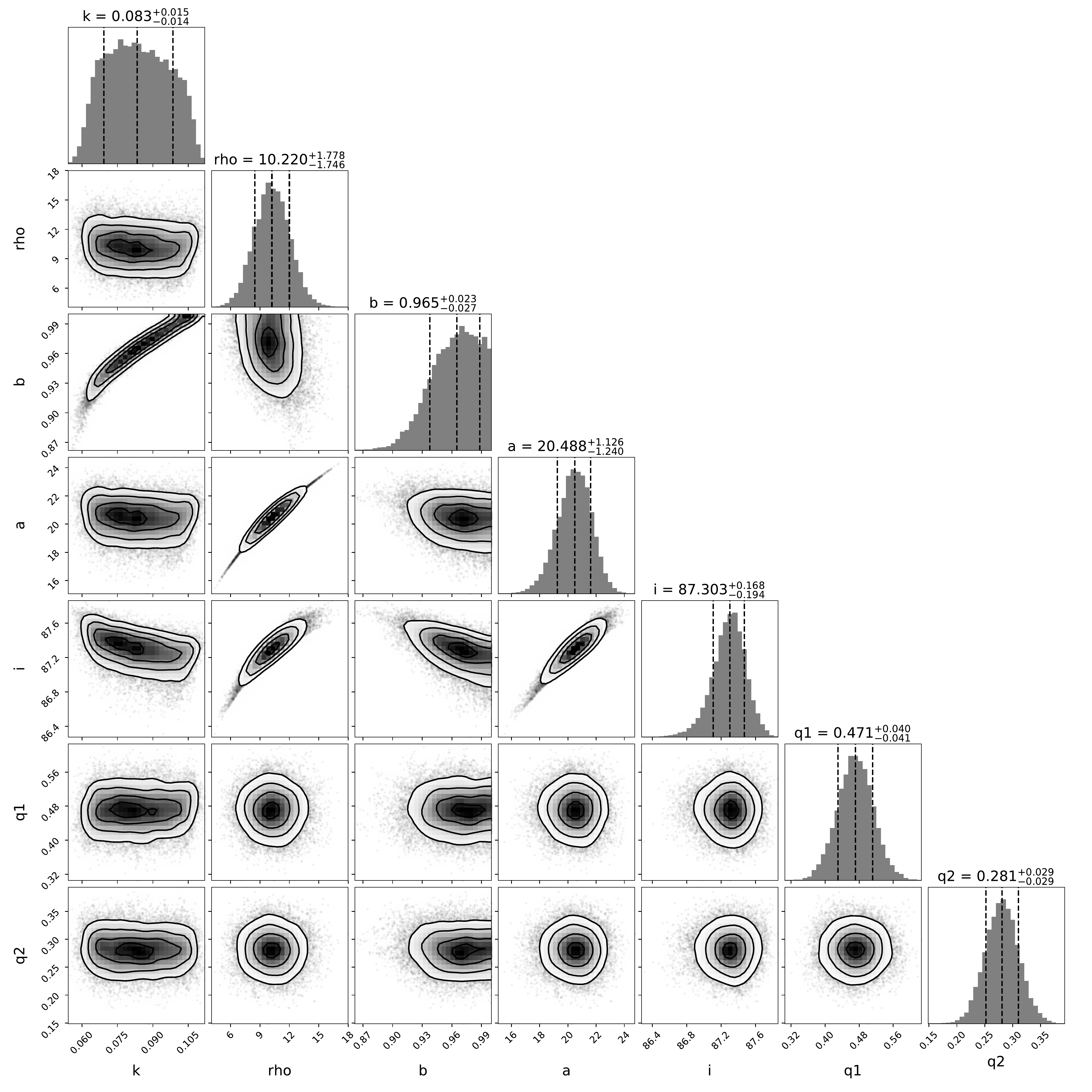}
\caption{Corner plot of the posterior distributions of the transit parameters for K2-146 c. The values above each column are the means of the posterior distributions with their respective 1-$\sigma$ uncertainties. Note that the impact parameter is not well-constrained. \label{fig:lc_cornerplot_c}}
\end{center}
\end{figure}

\begin{figure}[htbp!]
\begin{center}
\includegraphics[width=0.9\textwidth,trim=0.0cm 5.5cm 0.0cm 5.0cm,clip]{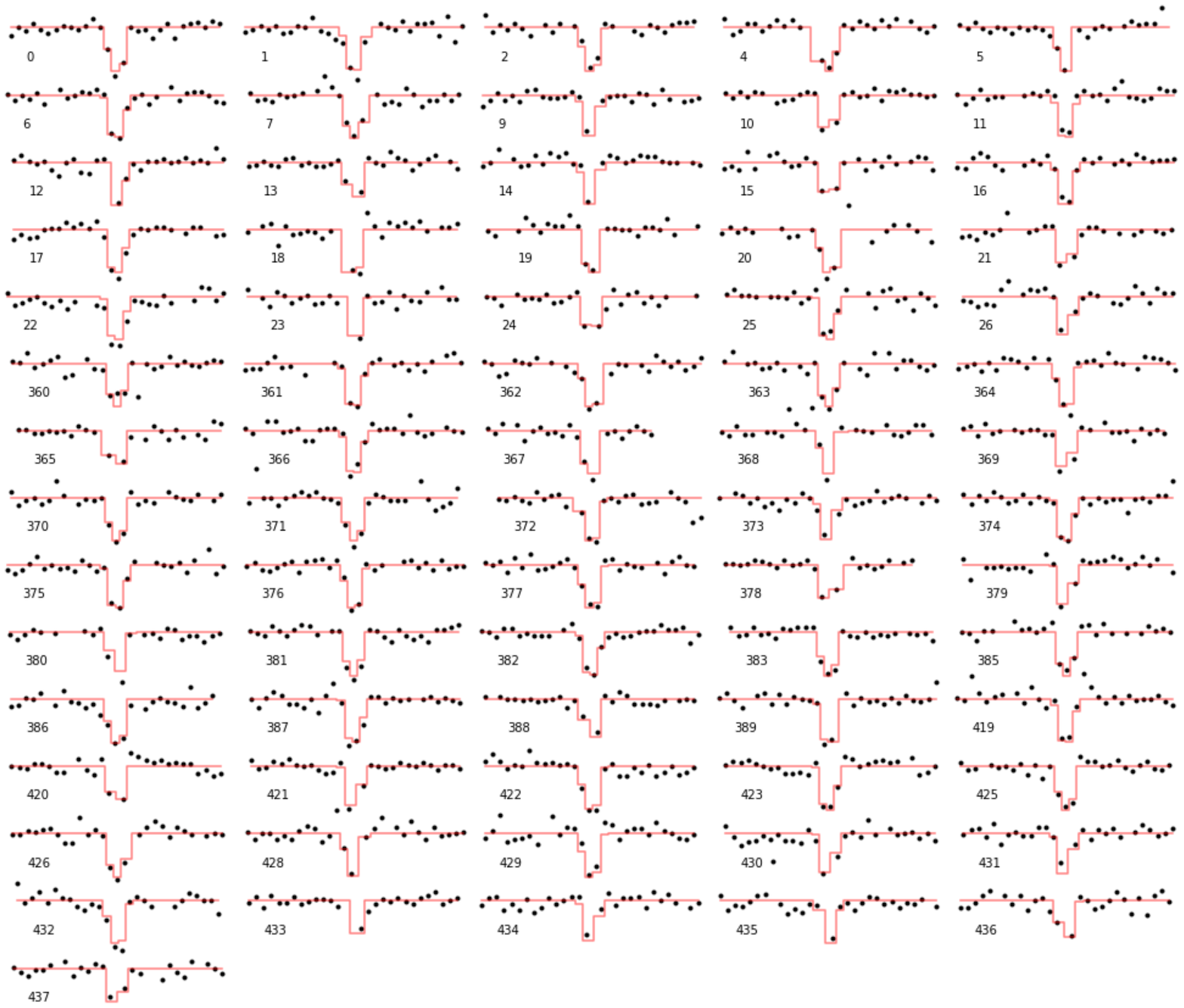}
\caption{Individual transits of K2-146 b fitted in the \texttt{PyTV} transit parameter analysis. The \textit{K2} data are denoted by the black points and the red solid lines are the best-fit transit models. The numbers in the bottom left corner of each transit plot correspond to the transit epoch number labelled in Figure \ref{fig:K2_lightcurve}. \label{fig:lc_individual_b}}
\end{center}
\end{figure}

\begin{figure}[htbp!]
\begin{center}
\includegraphics[width=0.9\textwidth,trim=0.0cm 15.5cm 0.0cm 5.0cm,clip]{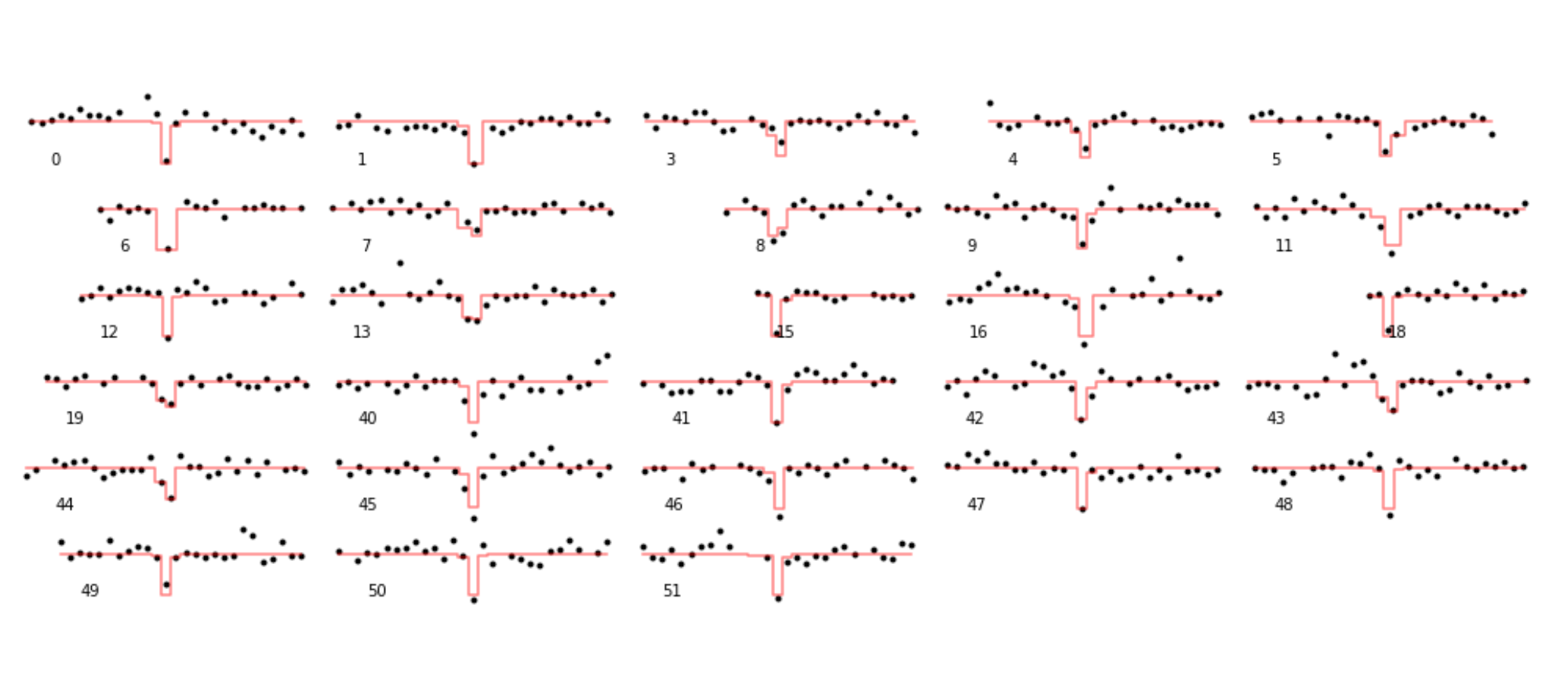}
\caption{Individual transits of K2-146 c fitted in the \texttt{PyTV} transit parameter analysis. Notations as in Figure \ref{fig:lc_individual_c}. \label{fig:lc_individual_c}}
\end{center}
\end{figure}

    \begin{table*}[htbp!]
\centering
\caption{Transit parameters of K2-146 b and K2-146 c derived from \texttt{PyTV} analysis and stacked transit analysis. The reported posterior values reported are the median and 1-$\sigma$ uncertainties. \label{tab:transitparams}}
\begin{tabular}{lcccc}
\hline
 Parameter          & \multicolumn{2}{c}{\texttt{PyTV}} &\multicolumn{2}{c}{Stacked transit analysis}        \\
\hline 
                       & K2-146 b              & K2-146 c              &  K2-146 b  & K2-146 c  \\
\hline 
P [day]                & $2.65044 \pm 0.00007$ & $3.98974 \pm 0.00069$ &  -  & -  \\
$\rm R_p/R_s$          & $0.057   \pm 0.001$   & $0.084   \pm 0.013$   &  $0.0527 \pm 0.0004$  &  $0.0604 \pm 0.0046$ \\
b                      & $0.27    \pm 0.19$    & $0.96    \pm 0.02$    &  $0.03 \pm 0.03$  & $0.81 \pm 0.03$  \\
$\rm a/R_s$           & $15.06   \pm 1.16$    & $20.44   \pm 1.19$    &  $15.51 \pm 0.03$  &  $20.38 \pm 0.07$ \\
$i$ [\degree]            & $88.9    \pm 0.9$     & $87.3    \pm 0.2$     &  $89.9 \pm 0.11$  &  $87.7 \pm 0.09$ \\
$\rm R_p$ [\Rearth] & $2.18    \pm 0.22$    & $3.16    \pm 0.59$    &  $2.01 \pm 0.02$  & $2.31 \pm 0.18$  \\
a [AU]                 & $0.0246  \pm 0.0031$  & $0.0332  \pm 0.0039$  &  $0.0252 \pm 0.0166$  &  $0.0332 \pm 0.0001$ \\
\hline                                                                          \\      
\end{tabular}
\end{table*}

\newpage

\end{document}